\documentclass[%
 reprint,
 superscriptaddress,
 amsmath,amssymb,
 aps,
]{revtex4-2}

\usepackage{graphicx}
\usepackage{dcolumn}
\usepackage{bm}
\usepackage{hyperref}
\hypersetup{
    colorlinks=true,
    linkcolor=blue,
    citecolor=blue,
    filecolor=magenta,      
    urlcolor=blue,
    pdfpagemode=FullScreen,
    }
\usepackage{physics}
\usepackage{float}
\usepackage[dvipsnames]{xcolor}
\usepackage{mathtools}
\usepackage{siunitx}
\usepackage{algorithm}
\usepackage{algpseudocode}
\usepackage{blkarray}

\usepackage[english]{babel}
\usepackage[autostyle, english = american]{csquotes}
\MakeOuterQuote{"}
\begin{document}
\preprint{APS/123-QED}

\title{Implementation of a quantum addressable router using superconducting qubits}

\author{Connie Miao}
\thanks{These authors contributed equally to this work}
\email{\newline
sebleger@stanford.edu}
\affiliation{Departments of Physics and Applied Physics, Stanford University, Stanford, CA 94305, USA}

\author{Sébastien Léger}
\thanks{These authors contributed equally to this work}
\email{\newline
sebleger@stanford.edu}
\affiliation{Departments of Physics and Applied Physics, Stanford University, Stanford, CA 94305, USA}
\author{Ziqian Li}
\affiliation{Departments of Physics and Applied Physics, Stanford University, Stanford, CA 94305, USA}

\author{Gideon Lee}
\affiliation{Pritzker School of Molecular Engineering, University of Chicago, Chicago, IL 60637, USA}

\author{Liang Jiang}
\affiliation{Pritzker School of Molecular Engineering, University of Chicago, Chicago, IL 60637, USA}

\author{David I. Schuster}
\affiliation{Departments of Physics and Applied Physics, Stanford University, Stanford, CA 94305, USA}

\date{\today}


\newcommand{\leakageRbQOneQTwo}{1.88\pm 0.16\%} 
\newcommand{\leakageRbQOneQThree}{0.83\pm 0.01\%} 

\newcommand{\leakageRbQOneQTwoWrong}{0.88 \pm 0.22 \%}
\newcommand{\leakageRbQOneQThreeWrong}{1.82 \pm 0.02 \%}

\newcommand{\errRbQOneQTwo}{3.05 \pm  0.11\%}
\newcommand{\errRbQOneQThree}{1.36 \pm 0.01\%}

\newcommand{\errRbQOneQTwoWrong}{3.86 \pm 0.48 \%}
\newcommand{\errRbQOneQThreeWrong}{2.87 \pm 0.02 \%}


\newcommand{\errIrbQOneQTwo}{2.24 \pm 0.13\%}
\newcommand{\errIrbQOneQThree}{1.38 \pm 0.02\%}

\newcommand{\fidIrbQOneQTwo}{97.77 \pm 0.13\%}
\newcommand{\fidIrbQOneQThree}{98.62 \pm 0.02\%}

\newcommand{\leakageRbQOneQTwoX}{1.32 \pm 0.08\%} 
\newcommand{\leakageRbQOneQThreeX}{0.85\pm 0.02\%} 

\newcommand{\fidProtocolTomo}{95.3\%}
\newcommand{\errProtocolSim}{2.7\%}
\newcommand{\errSPAMtomo}{1.7\%}

\newcommand{\errAmpCrCi}{1.6\pm 0.9\%}
\newcommand{\errAmpCrQi}{2.6\pm 1.3\%}
\newcommand{\errAmpQrCi}{2.4\pm 0.9\%}
\newcommand{\errAmpQrQi}{3.3\pm 0.4\%}

\newcommand{\fidErrAmp}{97.5\pm 0.5\%}

\newcommand{\errSimCrCi}{2.0\%}
\newcommand{\errSimCrQi}{2.7\%}
\newcommand{\errSimQrCi}{2.7\%}
\newcommand{\errSimQrQi}{3.2\%}

\newcommand{\badStateFifteenSwapQTwo}{3.6\%}
\newcommand{\badStatePerSwapQTwo}{0.3\%}
\newcommand{\badStateFifteenSwapQThree}{2.2\%}
\newcommand{\badStatePerSwapQThree}{0.1\%}

\begin{abstract}


The implementation of a quantum router capable of performing both quantum signal routing and quantum addressing (a Q$^2$-router) represents a key step toward building quantum networks and quantum random access memories. We realize a Q$^2$-router that uses fixed-frequency transmon qubits to implement a routing protocol based on two native controlled-iSWAP gates. These gates leverage a large ZZ interaction to selectively route information according to a quantum address. We find an estimated average routing fidelity of $\fidProtocolTomo$, with errors arising primarily from decoherence or state preparation and measurement. We present a comprehensive calibration and characterization of both the c-iSWAP gates and the overall routing protocol through randomized benchmarking techniques and state tomography.

\end{abstract}

\maketitle

\renewcommand{\figurename}{FIG.}

\section{Introduction}

Routing information is crucial for classical processing and is even more essential in quantum systems due to the inherent constraint that quantum states cannot be duplicated~\cite{buzek_quantum_1996}.
The most general form of routing is achievable with a Q$^2$-router, which allows both the data and address to be quantum mechanical~\cite{lemr_resource-efficient_2013}. Classical routing of quantum states plays a key role in quantum networking~\cite{cirac_quantum_1997, kimble_quantum_2008}. Similarly, the ability to direct information to superpositions of paths via quantum addresses enables efficient quantum random access memory (QRAM)~\cite{giovannetti_architectures_2008, giovannetti_quantum_2008, xu_systems_2023}, which can aid algorithms such as Grover's search~\cite{grover_fast_1996}, quantum matrix inversion~\cite{harrow_quantum_2009}, quantum deep learning~\cite{wiebe_quantum_2015}, and quantum simulation~\cite{berry_qubitization_2019, babbush_encoding_2018}. Additionally, quantum addressing itself offers applications in quantum communication~\cite{miguel-ramiro_genuine_2021, chiribella_quantum_2019, bartkiewicz_using_2014}, quantum calibration~\cite{barenco_stabilization_1997, buhrman_quantum_2001}, data compression~\cite{liu_data_2023}, and error filtration~\cite{lee_error_2023, miguel-ramiro_superposed_2023}.

Numerous platforms have experimentally explored classical routing of quantum states, including routing between qubits~\cite{wu_modular_2024, zhou_realizing_2023, naik_random_2017} and the directional emission of microwave~\cite{pechal_superconducting_2016, kannan_-demand_2023-1, almanakly_deterministic_2024, hoi_demonstration_2011} and optical~\cite{sollner_deterministic_2015, lodahl_chiral_2017, coles_chirality_2016} photons. However, the classical routers in these studies do not possess an internal addressing mechanism, as the routing direction must be encoded externally. This encoding is typically achieved, for instance, by applying different pulses corresponding to the routing direction or by altering the design of the sample. On the other hand, routing a state based on a quantum address has received various theoretical proposals based on photonics~\cite{chen_scalable_2021}, Rydberg atoms~\cite{hong_robust_2012}, superconducting circuits~\cite{weiss_quantum_2024-1, hann_hardware-efficient_2019}, and hybrid architectures~\cite{hann_hardware-efficient_2019, wang_quantum_2024}. Alongside these proposals, several groups~\cite{hu_native_2023, chapman_high--off-ratio_2023, warren_extensive_2023-1} have implemented three-qubit gates that could be used as building blocks for a quantum addressing operation. A Q$^2$-router has been implemented on a photonic platform~\cite{yuan_experimental_2015}, but its dependence on non-deterministic post-selection hinders its applicability. As a result, a fully deterministic Q$^2$-router has yet to be demonstrated. 

In this work, we use superconducting circuits to realize a Q$^2$-router able to handle both a quantum signal and a quantum address. For minimal overhead, we employ fixed-frequency transmon qubits. The routing protocol, inspired by previous theoretical proposals~\cite{hann_hardware-efficient_2019}, consists of two controlled-iSWAP (c-iSWAP) gates. Each of these gates is natively implemented with an $\ket{eg}$ to $\ket{gf}$ two-qubit sideband gate and uses a large ZZ interaction to control the iSWAPs. The fidelities of these gates are estimated to be $\fidIrbQOneQTwo$ and $\fidIrbQOneQThree$ within the valid subspace. The routing protocol has an average fidelity of $\fidProtocolTomo$ and we estimate that qubit decoherence and state preparation and measurement are the main sources of error. In Sec. \ref{sec:layout_protocol} we present our router design and protocol. In Sec. \ref{sec:calib} we explain how the c-iSWAPs used in the protocol are calibrated and we benchmark them individually using randomized benchmarking and error amplification. Finally, in Sec. \ref{sec:protocol_benchmarking} we estimate the fidelity of the routing protocol using state tomography and error amplification. 

\section{Router layout and protocol \label{sec:layout_protocol}}

\begin{figure*}[!htb]
\centering
\includegraphics[width=1\textwidth]{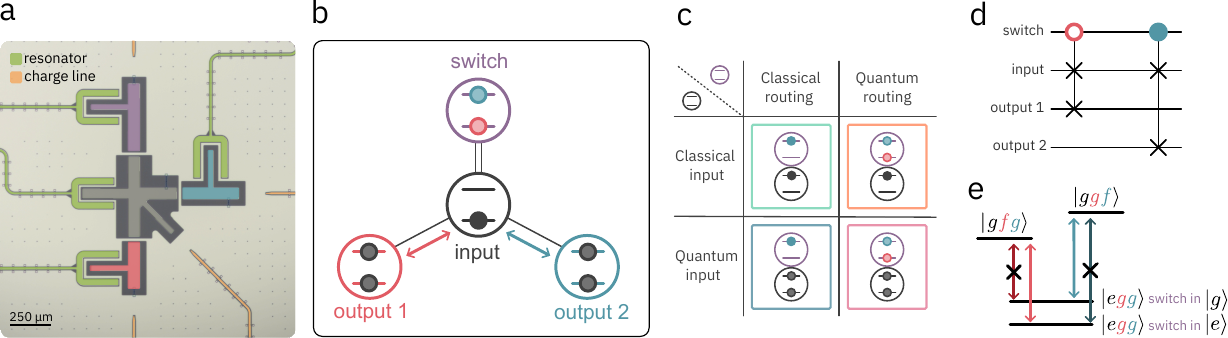}
\caption{\label{fig1}\textbf{Device and $\mathrm{Q}^2$-router concept.} \textbf{(a)} False colored optical microscope image of the $\mathrm{Q}^2$ router device with four transmon qubits (switch (purple), input (dark grey), output 1 (red), and output 2 (blue)) and their readout resonators (green) and charge drive lines (yellow). \textbf{(b)} Schematic of our $\mathrm{Q}^2$-router concept in a possible three-qubit entangled final state after one routing operation. Shaded state circles indicate superposition states while solid circles indicate classical states. Lines between pairs of qubits indicate strong couplings; the switch and input qubit are designed to be particularly strongly ZZ coupled (5.4\,MHz). The red and blue arrows indicate the pairs of qubit between which we drive two-qubit swaps to implement the protocol. \textbf{(c)} Breakdown of the possible modes of operation of a general quantum router depending on whether the input and switch (routing) is classical or quantum. The $\mathrm{Q}^2$-router is capable of performing the most general form of routing where both the input and routing are quantum. \textbf{(d)} Quantum circuit describing the protocol: 1. iSWAP between input and output 1, controlled on switch in $\ket{g}$; 2. iSWAP between input and output 2, controlled on switch in $\ket{e}$. \textbf{(e)} Physical implementation of the routing protocol using native $\ket{eg}$ to $\ket{gf}$ swaps with levels labeled for input and output states. The controlled aspect of each swap is implemented via the large ZZ interaction between the switch and input, which causes the swap to be blocked when the switch is in the wrong initial state.}
\end{figure*}


Our $\mathrm{Q}^2$-router is based on four fixed-frequency, grounded transmon qubits with fixed capacitive couplings. An optical image of the sample is shown in Fig. \ref{fig1}(a) with the connectivity labeled in Fig. \ref{fig1}(b). The circuit can be used in any of the four classical/quantum modes of information routing shown in Fig. \ref{fig1}(c).

To enable fast two-qubit gates, we have strong couplings on the order of $g/2\pi\sim50$\,MHz between the input and the switch as well as between the input and each of the two output qubits. The key ingredient to enable native c-iSWAPs is a relatively strong ZZ interaction between the input and switch ($\zeta_{\mathrm{SI}}/2\pi=5.4$\,MHz). Each qubit is also capacitively coupled to its own charge line for control and its own resonator, which enables four-qubit simultaneous readout. This sample was fabricated and packaged by the MIT SQUILL foundry and the measured qubit parameters and lifetimes are reported in Table \ref{tab:qubits} in Appendix \ref{app:param}. 

In our convention, the input state is routed to output 1 (2) if the switch is in $\ket{g}$ ($\ket{e}$), or to a superposition of the outputs with the relative amplitudes given by the switch amplitudes in $\ket{g}$ or $\ket{e}$. This routing operation $U_\mathrm{route}$ can be written as
\begin{equation*}
\begin{aligned}
    \ket{\psi}_\mathrm{I} \Bigl(\alpha\ket{0}  + \beta\ket{1}&\Bigr)_\mathrm{S} \ket{0}_\mathrm{O_1}\ket{0}_\mathrm{O_2}\\
    &\downarrow{U_{\mathrm{route}}}\\
    \ket{0}_\mathrm{I} \Bigl(\alpha\ket{0}_\mathrm{S}\ket{\psi}_\mathrm{O_1}\ket{0}_\mathrm{O_2} & + \beta\ket{1}_\mathrm{S}\ket{0}_\mathrm{O_1}\ket{\psi}_\mathrm{O_2}\Bigr),
\end{aligned}
\end{equation*}
where I, S, O$_1$ and O$_2$ represent the input, switch, output 1, and output 2 qubits respectively. To perform the routing protocol, we tile one c-iSWAP, which swaps the input and output 1 controlled on the switch in $\ket{g}$, and one c-iSWAP, which swaps the input and output 2 controlled on the switch in $\ket{e}$ (Fig. \ref{fig1}(d)). Valid router states and operations are a subset of the full four-qubit state and control space. Most importantly, the router module should have at most one total photon shared between the input and outputs at all times, with the routing protocol swapping this photon between the input and the outputs. Thus, for simplicity, we always initialize the outputs in their ground states and so we expect the input to always be in its ground state at the end of the protocol. Because we only need the protocol to operate correctly within this restricted subspace, within the protocol any physical Z rotations can be compensated with a virtual one (Appendix \ref{app:virtual_Z}).

We realize our c-iSWAP gate by driving the $\ket{eg}$ to $\ket{gf}$ transition between the input and output 1 (2) with a sideband pulse applied on output 1 (2)~\cite{zeytinoglu_microwave-induced_2015} during a long enough time to perform a $\sqrt{\mathrm{iSWAP}}$. We apply two $\sqrt{\mathrm{iSWAP}}$ pulses to implement an iSWAP between the input and output qubits. To make the iSWAP conditional, we leverage the strong ZZ interaction between the input and switch. Because of this interaction, the input qubit, and therefore the $\ket{eg}-\ket{gf}$ frequencies, depends on the state of the switch (Fig. \ref{fig1}(e)). With an appropriate gate calibration, we can thus ensure that population is only transferred when the switch is in the desired state. We note that combined with the same ZZ-based principle to implement the control, other two-qubit sideband schemes could be used instead of $\ket{eg}-\ket{gf}$. We choose the $\ket{eg}-\ket{gf}$ interaction because it is a single photon transition, making it fast and simple to implement~\cite{zeytinoglu_microwave-induced_2015, egger_entanglement_2019}, especially given our relatively far-detuned qubit frequencies. With our protocol the outputs are encoded in the $g$-$f$ subspace as opposed to the more standard $g$-$e$, which is a valid choice for either networking or QRAM applications (Appendix \ref{app:scaling}).

\section{c-iSWAP calibration and characterization \label{sec:calib}}

\subsection{Calibration}\label{sec:swap_calib_leakage}

\begin{figure}
\centering
\includegraphics[width=\columnwidth]{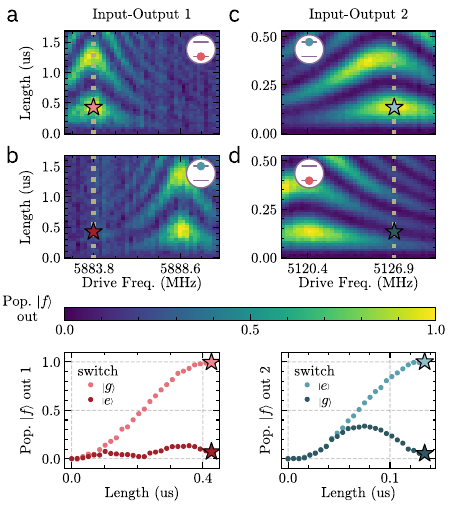}
\caption{\label{fig2}\textbf{c-iSWAP calibration.} \textbf{(a, c)} Rabi oscillations for each swap measuring the target output $\ket{f}$ population, sweeping pulse length vs. drive frequency when the switch is initialized in $\ket{S}$, the state to allow the swap. The calibrated pulse time is marked with a light red or blue star. \textbf{(b, d)} Same experiment as (a, c) but initializing the switch in $\ket{\bar{S}}$, the state to block the swap. With this initialization, there is a minimum of population transfer at the calibrated pulse time (dark red or blue star). \textbf{(e, f)} Slices through the Rabi oscillations in previous panels at the calibrated swap frequency (dashed lines). At the calibrated length (stars), we observe an iSWAP ($4\pi$ rotation) when the switch is in $\ket{S}$ ($\ket{\bar{S}}$).}
\end{figure}
 
To achieve high-fidelity c-iSWAPs, we aim for two goals. First, each sideband pulse should implement a high-fidelity iSWAP. Second, the two pulses should be selective, performing the swap only when the switch is in the correct state $\ket{S}$ for the iSWAP to proceed, where $\ket{S}=\ket{g}$ for input to output 1 or $\ket{S}=\ket{e}$ for input to output 2.

To implement the iSWAP, we need to calibrate the drive frequency, amplitude, and gate time. These parameters are interdependent because the qubits are AC-stark shifted during the drive~\cite{zeytinoglu_microwave-induced_2015}, so we first establish a relationship between them through an initial calibration for different possible pulse lengths (Appendix \ref{app:swap_calib}). We choose pulse parameters to maximize contrast given the finite $\zeta_\mathrm{SI}/2\pi = 5.4$\,MHz shift in the sideband, such that the desired transition is driven strongly at rate $g_\mathrm{eg-gf}$ while the undesired transition is weakly driven at $\Omega$. It is possible to ensure no state transfer in the blocking state despite still having weak driving if we pick $\Omega = \sqrt{g_\mathrm{eg-gf}^2 + \zeta_\mathrm{SI}^2}$ to be $2n\,g_\mathrm{eg-gf}$ where $n$ is an integer \cite{tsunoda_error-detectable_2023}. The calibration of the optimal point is explained in Appendix \ref{app:swap_calib}.

In Fig. \ref{fig2}(a) (\ref{fig2}(c)) we show the resulting Rabi oscillation between the input and output 1 (2) versus pulse length and drive frequency at the calibrated drive amplitude when the switch qubit is in $\ket{S}$.
In Fig. \ref{fig2}(b) (\ref{fig2}(d)), we prepare the switch in the blocking state $\ket{\bar{S}}=\ket{e}$ ($\ket{\bar{S}}=\ket{g}$) and again measure the Rabi oscillation between the input and output 1 (2) versus pulse length and drive frequency. We see that the resonant frequency is shifted and that the previously calibrated point now corresponds to a minimum of population in the output qubit. In Figs. \ref{fig2}(e, f), we show slices through the Rabi oscillations from the previous panels at the selected optimal frequency, initializing the switch in either $\ket{S}$ or $\ket{\bar{S}}$.
The pulse times are, respectively, 426\,ns and 134\,ns, corresponding to $n=2$ and $n=1$ for output 1 and 2 (see Appendices \ref{app:swap_calib}, \ref{app:rb_other_minima} for more details).



\begin{figure*}[hbt]
\centering
\includegraphics[width=1.0\textwidth]{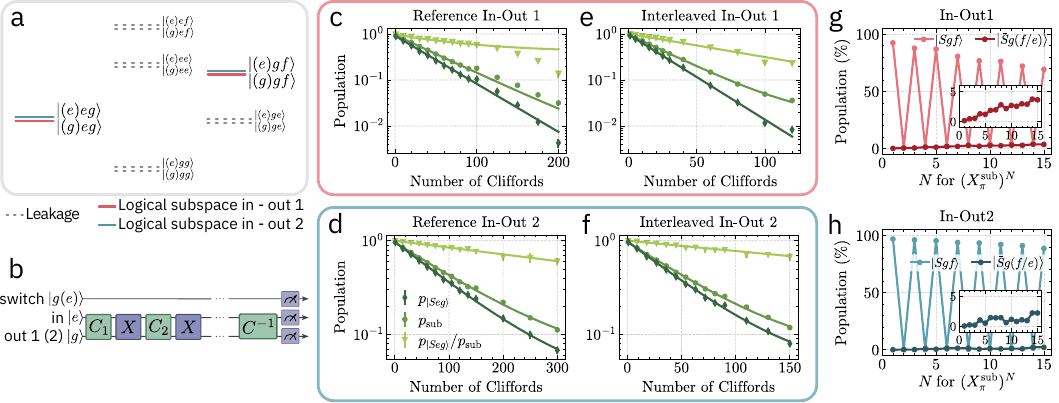}
\caption{\label{fig3} \textbf{Subspace randomized benchmarking of c-iSWAPs.} \textbf{(a)} Level diagram of states probed for each randomized benchmarking (RB) experiment on the switch, input and relevant output. We treat $\ket{Seg}$-$\ket{Sgf}$ as the logical two-level subspace and the Clifford group is defined relative to this subspace. Population measured in any other states is considered leakage. \textbf{(b)} Gate sequence used to implement interleaved RB. $C_i$ is the $i^\mathrm{th}$ randomly selected Clifford gate while $C^{-1}$ is the inversion Clifford. For the reference RB, the interleaved $X$ is removed. \textbf{(c, d)} Reference RB measured population vs. number of Clifford gates. $p_{\ket{Seg}}$ is the survival probability, $p_\mathrm{sub} = p_{\ket{Seg}} + p_{\ket{Sgf}}$ is the probability to stay in the logical subspace and $p_{\ket{Seg}}/p_\mathrm{sub}$ is the survival probability post-selected for the counts that are still in the subspace. \textbf{(e, f)} Interleaved RB measured population vs. number of Clifford gates.
\textbf{(g, h)} Error amplification of the c-iSWAP when the switch state is in either $\ket{S}$ (light red, light blue) or $\ket{\bar{S}}$ (dark red, dark blue), measuring the population in $\ket{Sgf}$ or in $\ket{\bar{S}g(f/e)}$ respectively. After 15 gates, we find $\badStateFifteenSwapQTwo$ transfer ($\badStatePerSwapQTwo$ per c-iSWAP) into output 1 and $\badStateFifteenSwapQThree$ transfer ($\badStatePerSwapQThree$ per c-iSWAP) into output 2 when the switch is in $\ket{\bar{S}}$.}
\end{figure*}


\subsection{Benchmarking individual c-iSWAPs}

Before benchmarking the full router protocol we first quantify the errors in the individual c-iSWAPs. We treat the subspace $\ket{Seg}-\ket{Sgf}$ as a single dual-rail logical qubit on which we perform reference and interleaved randomized benchmarking (RB)~\cite{magesan_characterizing_2012, knill_randomized_2008, lu_high-fidelity_2023, magesan_efficient_2012}. Within the logical subspace, the sideband is described by the single qubit rotation unitary  
\begin{equation*}
 U^\mathrm{sub}(\varphi) = \exp\left[-i\frac{\pi}{4}\left(e^{i\varphi}\ketbra{Seg}{Sgf} + e^{-i\varphi}\ketbra{Seg}{Sgf}\right)\right].
\end{equation*}
From $U^\mathrm{sub}$, we define $X^\mathrm{sub}_{\pi/2} \equiv U^\mathrm{sub}(0)$ and $Y^\mathrm{sub}_{\pi/2} \equiv U^\mathrm{sub}(\pi/2)$. The Clifford set is constructed from $\{I, \pm X^\mathrm{sub}_{\pi/2}, \pm Y^\mathrm{sub}_{\pi/2} \}$. To perform the reference RB (Figs. \ref{fig3}(c, d)), we prepare the switch, input, and target output in $|Seg\rangle$ and apply a sequence of gates selected randomly from the Clifford group. For the interleaved RB (Figs. \ref{fig3}(e, f)), we add an extra gate that we wish to characterize after each randomly selected Clifford gate in the sequence (Fig. \ref{fig2}(b))~\cite{magesan_efficient_2012}. In both cases we then apply the inversion gate and measure the switch, input, and target output simultaneously to determine the probability of these qubits having returned successfully to their initial states. The full RB experimental details and data analysis procedures are described in Appendix \ref{app:rb}. 

An advantage of treating the c-iSWAP as a conditional dual-rail single qubit is that we can distinguish errors that leave the subspace (Fig. \ref{fig3}(a)) by measuring the populations of all relevant qubit levels $\{ \ket{g/e}\otimes\ket{g/e}\otimes\ket{g/e/f}\}$. We estimate the probability of remaining in the logical subspace $p_\mathrm{sub}=p_{\ket{Seg}} + p_{\ket{Sgf}}$ at the end of the sequence, as shown in Figs. \ref{fig3}(c-f). From fitting $p_\mathrm{sub}$ (Eq. \ref{eq:prob_subspace_fit}) in the reference RB, we can estimate a photon loss per Clifford for each sideband of
$L_\mathrm{IO_1}=\leakageRbQOneQTwo$ ($L_\mathrm{IO_2}=\leakageRbQOneQThree$)
for input-output 1 (2), where error bars represent 95\% confidence from the fitting error. Interestingly, we see that the survival probability $p_{\ket{Seg}}$ closely matches the subspace survival $p_\mathrm{sub}$, indicating that the gate error is dominated by photon loss. The remaining logical error is estimated from the survival probability post-selected on the probability to stay in the logical subspace, defined as $p_\mathrm{survival}=p_{\ket{Seg}}/p_\mathrm{sub}$, also shown in Figs. \ref{fig3}(c-f). Fitting $p_\mathrm{survival}$ from the reference RB with an exponential (Eq. \ref{eq:prob_survival_fit}) gives the total error taking into account the photon loss error as
$\epsilon_\mathrm{IO_1}=\errRbQOneQTwo$ ($\epsilon_\mathrm{IO_2}=\errRbQOneQThree$) per Clifford.
Notably, the reference RB for input-output 1 deviates from an exponential for circuits with over 150 gates, likely due to the small $p_\mathrm{sub}$ above this depth. We limit exponential fitting to depths $<150$ to make this deviation apparent, though fitting up to 200 gates yields the same error estimate.

For the routing protocol, the only necessary gate is $X^\mathrm{sub}_\pi$ between the input and each output. To measure the fidelity of this gate specifically we perform interleaved RB. Following the same experimental and fitting procedure as in the reference RB, we estimate both the photon loss contribution and the total error per gate, with the results shown in Figs. \ref{fig3}(e, f). For the iSWAP between input and output 1 (2), we estimate the average total error per $X^\mathrm{sub}_\pi$ to be
$\epsilon^X_\mathrm{IO_1}=\errIrbQOneQTwo$ ($\epsilon^X_\mathrm{IO_2} =\errIrbQOneQThree$), with photon loss rate per $X^\mathrm{sub}_\pi$ $L^X_\mathrm{IO_1}=\leakageRbQOneQTwoX$ ($L^X_\mathrm{IO_2}=\leakageRbQOneQThreeX$) (see Appendix \ref{app:rb} for more details). The gate error for input-output 2 is about half that of the gate with output 1. This difference is mainly due to the input-output 2 gate being shorter and therefore less sensitive to decoherence. These fidelities are comparable with other c-SWAP-like gates implemented with superconducting or bosonic qubits~\cite{warren_extensive_2023-1, chapman_high--off-ratio_2023, hu_native_2023}. We note that because our protocol assumes there is at most one excitation shared between the input and output qubits, we do not measure the fidelity of our c-iSWAP on $\ket{ef}_\mathrm{IO}$.


After benchmarking each iSWAP, we confirm proper control by the switch, using error amplification to highlight amplitude deviations from a perfect $2\pi n$ rotation when the switch is in the blocking state $\ket{\bar{S}}$. We prepare the switch in $\ket{\bar{S}}$ and measure the population in both $\ket{\bar{S}gf}$ and $\ket{\bar{S}ge}$ after applying $N=15$ pulses (Figs. \ref{fig3}(g, h)). We include both of these states to account for population that may have transferred to the output and decayed over the course of the error amplification. We use this method instead of RB because it coherently adds unintended small rotations, making it a more sensitive measure of the contrast. With this procedure, we find on average $\badStatePerSwapQTwo$ ($\badStatePerSwapQThree$) undesired transfer per c-iSWAP between input and output 1 (2). To demonstrate the c-iSWAP on-off ratio we also initialize the switch in $\ket{S}$ and observe decaying oscillations as a function of $N$. 


With their simplicity, high on-off ratio, and high fidelity in the non-blocking state, these gates meet all the key criteria for integration into the routing protocol.

\section{Protocol benchmarking \label{sec:protocol_benchmarking}}

\begin{figure*}[!htb]
\centering
\includegraphics[width=\textwidth]{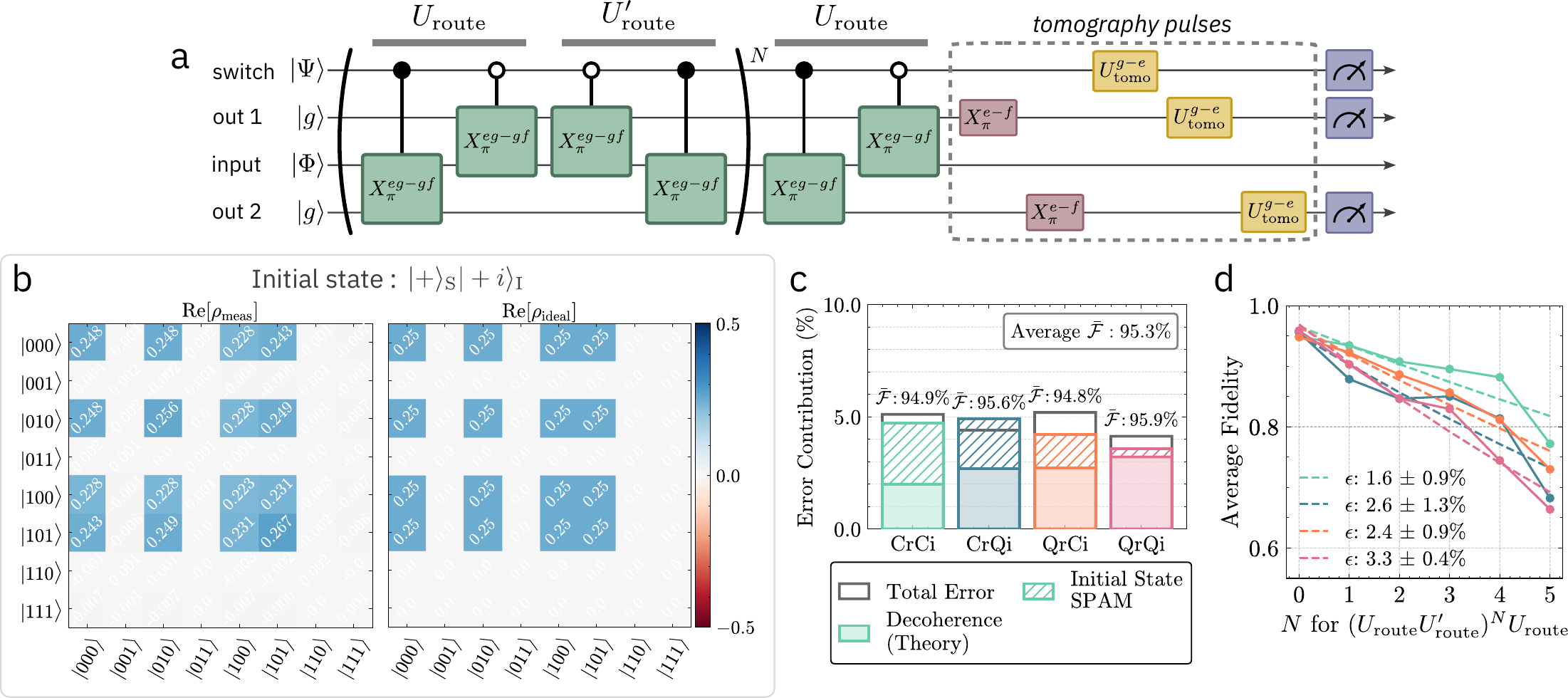}
\caption{
\label{fig4}
\textbf{Protocol fidelity from state tomography.}
\textbf{(a)} Pulse diagram for tomography experiments. Tomography pulses on the $g$-$e$ basis $U_\mathrm{tomo}^{g-e}$ are applied based on the measurement basis. A single protocol corresponds to the pulse diagram with $N=0$. \textbf{(b)} Left (right) panel: real component of the measured (ideal) 3Q tomography on the switch and outputs 1 and 2 after one routing protocol $U_\mathrm{route}$ starting with the switch in $\ket{+}$ and the input in $\ket{+i}$. The measured density matrix has a fidelity of $\mathcal{F} = 96.4\%$ compared to the ideal final state. \textbf{(c)} Protocol fidelities and error contributions for the four switch and input configurations. C/Q denotes classical/quantum and r/i denotes routing/input. Each configuration is averaged over the four corresponding cardinal states. Total error and fidelities (gray) are calculated compared to the ideal state as in (b). Decoherence error (shaded fill) is estimated from master equation simulations using the measured qubit coherence times (Table \ref{tab:qubits}). State preparation error (hatched fill) is estimated by calculating fidelity with respect to the measured initial state on the switch and input instead of the ideal one. \textbf{(d)} To amplify error per routing operation and disentangle gate error from SPAM, we measure the fidelity decay of the final state versus the number of applied forward ($U_{\mathrm{route}}$) and backward ($U_{\mathrm{route}}'$) protocols. The error is estimated using Eq. \ref{eq:fid_decay} and aligns with the decoherence limited fidelity. 
}
\end{figure*}


To quantify the performance of the router we apply the full protocol $U_\text{route}$ and then perform three-qubit (3Q) state tomography on the switch, output 1, and output 2  (Fig. \ref{fig4}(a), experimental details in Appendix \ref{app:tomography_methods}). In all of our 3Q tomography, we skip measurement of the input, as it is expected to always be in the ground state at the end of $U_\text{route}$ regardless of the initialization. 
We select 16 cardinal states to initialize on the switch and input, constructed from the tensor product of classical ($\ket{0}$, $\ket{1}$) or superposition ($\ket{+}\equiv(\ket{0}+\ket{1})/\sqrt{2}$, $\ket{+i}\equiv(\ket{0}+i\ket{1})/\sqrt{2}$) states. We build the initial state preparation and tomography pulses using optimal control pulses to realize $g$-$e$ pulses that are robust against the strong ZZ coupling between the switch and input (Appendix \ref{app:single_qubit_gates}). Before applying the tomography pulses we bring the outputs back to the $g$-$e$ subspace using $\pi$ $e$-$f$ pulses. 

To characterize the performance of the protocol for each cardinal state initialization, we calculate the fidelity of the measured density matrix compared to the ideal final state. We show this comparison for the $\ket{+}_\mathrm{S}\ket{+i}_\mathrm{I}$ initialization in Fig. \ref{fig4}(b).
In Fig. \ref{fig4}(c) we show the average fidelity binned in each of the four configurations of the routing and input behavior (classical/quantum). The average fidelity over all 16 initializations is $\mathcal{\bar{F}} = \fidProtocolTomo$, where each fidelity is calculated as $\mathcal{F}=\bra{\psi_\mathrm{ideal}}\rho_\mathrm{meas}\ket{\psi_\mathrm{ideal}}$~\cite{nielsen_quantum_2010}, $\rho_\mathrm{meas}$ is the measured density matrix, and $\ket{\psi_\mathrm{ideal}}$ is the ideal state.
Next, we estimate the contributions to the measured error. Using a master equation simulation taking into account the coherence times of the $g$-$e$ transitions of all qubits and the $e$-$f$ transitions of the two outputs (Table \ref{tab:qubits}), we find an average estimated error of $\errProtocolSim$ due to decoherence. Another error source is the imperfect preparation of the initial states, with a large contribution due to ZZ. We estimate SPAM error by performing 2Q tomography on the initial states and sending the measured density matrices through the ideal routing unitary (Eq. \ref{eq:U_route}) to use as the reference for the fidelity calculation. The difference in fidelity from this calculation compared to the previous calculation that references the ideal final state is about $\errSPAMtomo$. Both the decoherence and SPAM errors are compiled in Fig. \ref{fig4}(c) and match well with the averaged measured error. The remaining discrepancy between the measured and total estimated errors likely arises from errors in reconstructing the density matrices in the 3Q and 2Q tomography (see Appendix \ref{app:err_breakdown} for a state-by-state comparison). 


Typically, benchmarking techniques such as randomized benchmarking, cycle benchmarking, or cross entropy benchmarking~\cite{magesan_characterizing_2012, boixo_characterizing_2018, erhard_characterizing_2019} can be used to quantify gate fidelity while being insensitive to SPAM errors. However, these approaches are difficult to apply to the full protocol due to the restricted subspace that is allowed to have population.
We instead estimate the routing fidelity by applying odd multiples of the protocol, where on every even iteration we apply the backward protocol $U_\mathrm{route}'$, defined as reversing the order of the two c-iSWAPs (Fig. \ref{fig4}(a)). The sequence is thus $(U_\text{route} U'_{\text{route}})^N U_\text{route}$ for variable $N$.
This method of error amplification is not guaranteed to follow an exponential decay as it is sensitive to coherent error, but it is a useful metric as a comparison to the RB. We obtain the error per protocol $\epsilon$ by fitting the measured fidelity averaged for each configuration $\mathcal{F}(N)$ to
\begin{equation}\label{eq:fid_decay}
    \mathcal{F}(N) = A (1-\epsilon)^{2N+1}.
\end{equation}
From this calculation, we find that the measured error rates are
$\errAmpCrCi$, $\errAmpCrQi$, $\errAmpQrCi$, and $\errAmpQrQi$
respectively for the CrCi, CrQi, QrCi, and QrQi configurations of classical/quantum routing/input, where the errors bars represent 95\% confidence from the fitting error, for an overall fidelity of $\fidErrAmp$. The error amplification results are shown in Fig. \ref{fig4}(d). These error rates closely match the decoherence-caused error rates estimated from simulation, $\errSimCrCi$, $\errSimCrQi$, $\errSimQrCi$, and $\errSimQrQi$ respectively for the four configurations, suggesting that the routing protocol is likely decoherence-limited. These errors can be understood intuitively: for all configurations, the $T_1$ decay of any qubit impacts the fidelity. For QrCi, both the $T_1$ decay and the dephasing of the switch degrade the fidelity. For CrQi, because the quantum input state is transferred to one of the output states, the dephasing of either the input or outputs (but not the switch) contributes to the error. For QrQi, dephasing on any qubit affects the protocol. This analysis explains why the error-amplified and simulated decay-limited fidelities are both smallest for the CrCi configuration, increase for QrCi and CrQi, and are largest for QrQi. 

Finally, we note that we cannot directly compare fidelities measured for individual c-iSWAPs via RB with those measured for the protocol via tomography. In the protocol, the two c-iSWAPs are applied sequentially, so the second c-iSWAP operates on an initial state that has undergone a wait time during which errors can occur, while no such wait time is relevant for the RB. In addition, the RB experiments always have one total excitation between the input and the output while the tomography-based benchmarking averages over cardinal states with no excitations. These initializations have higher fidelities due to reduced sensitivity to decoherence. Finally, since the RB experiments are never initialized in a superposition state, only tomography-based benchmarking probes the capability for quantum routing.

\section{Conclusion}

We have implemented a $\mathrm{Q}^2$-router with minimal overhead, capable of deterministically routing classical or quantum signals via classical or quantum addresses. The routing protocol uses two native c-iSWAP gates, enabled by an $\ket{eg}-\ket{gf}$ drive for population transfer and a strong ZZ interaction between the control qubit and one of the target qubits for selectivity. The measured routing fidelity is $\fidProtocolTomo$ from which about $\errSPAMtomo$ error is estimated to come from SPAM. The remaining error is compatible with our estimate of a decoherence-limited protocol. 


To enhance the future performance of $\mathrm{Q}^2$-routers, some straight forward solutions would be to either make the gate faster by increasing the ZZ interaction between the input and switch or improve our qubits' lifetimes. Also promising would be the implementation of an erasure scheme such as that used in dual-rail encodings. Specifically, in our protocol, states where any output qubit ends in $\ket{e}$ due to photon loss are invalid. Detecting and removing shots where this occurs could significantly improve the fidelity of the routing protocol.
Another method of improvement is to mitigate ZZ interactions by introducing tunable couplers, particularly between the control and input qubits. Reducing always-on stray ZZ interactions would improve single-qubit gate robustness and state preparation. It would also generalize our c-iSWAP gate to operate in cases with multiple total excitations in the input and outputs. Given our measured fidelities, with additional ZZ control we expect to be able to use our router as a building block for a few-memory layer QRAM (Appendix \ref{app:scaling}) or for a proof-of-concept quantum network.

\begin{acknowledgments}

C.M. and S.L. acknowledge support from the Air Force Office of Scientific Research (AFOSR) Multidisciplinary Research Program of the University Research Initiative (MURI) Grant No. W911NF2010177. C.M. acknowledges additional support from NSF GRFP Grant No. DGE-2146755. G.L. and L.J. acknowledge support from the ARO(W911NF-23-1-0077), ARO MURI (W911NF-21-1-0325), AFOSR MURI (FA9550-19-1-0399, FA9550-21-1-0209, FA9550-23-1-0338), DARPA (HR0011-24-9-0359, HR0011-24-9-0361), NSF (OMA-1936118, ERC-1941583, OMA-2137642, OSI-2326767, CCF-2312755), NTT Research, Packard Foundation (2020-71479), and the Marshall and Arlene Bennett Family Research Program. The device used in this work was fabricated and packaged by the Superconducting Qubits at Lincoln Laboratory (SQUILL) Foundry at MIT Lincoln Laboratory, with funding from the Laboratory for Physical Sciences (LPS) Qubit Collaboratory. We also thank Lincoln Labs for providing a Josephson traveling-wave parametric amplifier. We gratefully acknowledge Aaron Trowbridge for useful discussions on optimal control theory, Arnold Mong for insights on efficient MLE for tomography with ZZ correction, and all the members of the Schuster Lab for stimulating discussions and technical support. 

\end{acknowledgments}

\appendix
\renewcommand{\thefigure}{A\arabic{figure}}
\setcounter{figure}{0}

\section{Experimental setup \label{app:wiring}}

\begin{figure}
\centering
\includegraphics[width=1\columnwidth]{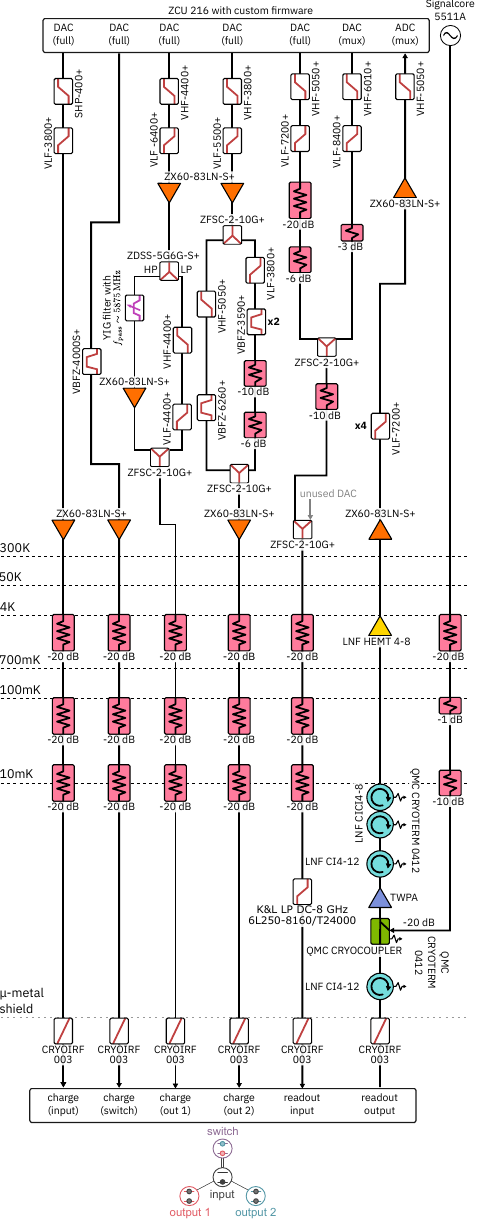}
\caption{\label{figures/wiring} \textbf{Room temperature and cryogenic wiring.}}
\end{figure}

The room temperature and cryogenic wiring setups are shown in Fig. \ref{figures/wiring}. The sample is packaged for us by the MIT SQUILL foundry in a gold plated OFHC copper housing~\cite{huang_microwave_2021} and heat sunk via OFHC copper to the mixing chamber (MXC) plate of a BlueFors LD400 dilution refrigerator, which is at approximately 10\,mK. The sample is entirely enclosed by an OFHC can that is coated on the inside with Berkeley black~\cite{persky_review_1999} to provide IR filtering then with two layers of \textmu-metal cans from MuShield to provide magnetic shielding. All input and output lines to the sample are filtered with CRYOIRF-003 IR filters placed inside the innermost can, and the readout input line is additionally filtered with a low pass filter from K\&L with a cut off around 8\,GHz.
To perform all of our qubit control and readout, we use a Xilinx Zynq UltraScale+ RFSoC ZCU216 with custom firmware made for us by the QICK collaboration~\cite{ding_experimental_2024, stefanazzi_qick_2022}. For qubit and sideband drives, we use four DAC generators ("full" type in QICK terminology) with sampling frequency $f_S=6389.760$\,MHz. We split and then recombine the output of the DACs for the two output qubits to enable separate filtering and amplification for the qubit and sideband drives. We use a YIG filter on the output 1 sideband drive to reduce populating the input qubit resonator, which is about 100\,MHz away from the sideband frequency. All four DACs for qubit control are coherently phase reset at the start of each experiment iteration.

On the readout side, we use an additional full DAC to drive the resonator tone on the switch and a multiplexed ("mux" type) DAC to drive the resonator tones on the input, output 1, and output 2. Both of these generators have $f_S=6881.280$\,MHz. To perform readout of all four resonator tones, we use a multiplexed ADC with $f_S=2457.600$\,MHz. Because we use a mixer-free setup, all readout tones are received in higher Nyquist zones of the mux ADC and are mapped back into the first Nyquist zone to perform the downconversion. 
We use a TWPA fabricated by Lincoln Labs that provides about 20-30\,dB of gain across the bandwidth of our resonator tone and we heat sink it via an OFHC copper finger to the MXC. 
We ensure that the environment surrounding the TWPA is impedance matched at both the readout frequencies and idler frequencies (9-11\,GHz) by using 4-12\,GHz circulators on either side of the TWPA. Two more 4-8\,GHz circulators provide additional isolation from the HEMT, which operates at 4-8\,GHz and provides about 35\,dB of gain. We use a SignalCore 5511A to provide a continuous pump tone for the TWPA. 
At room temperature, the readout signal is amplified by a ZX60-83LN-S+ amplifier from Minicircuits that is placed close to the fridge output flange. We use four VLF-7200+ low pass filters to attenuate the TWPA pump tone by approximately 20\,dB before amplifying again with another ZX60-83LN-S+. An additional VHF-5050+ high pass filter reduces noise from the lower Nyquist zones before the final readout acquisition.

\section{System parameters and simulation\label{app:param}}

In this section, we present the system model and the measured parameters. The system is composed of four capacitively coupled transmon qubits. Its Hamiltonian can be written as 
\begin{equation}
\begin{aligned}
H &= H_\mathrm{q} + H_\mathrm{c} \\
H_\mathrm{q} &= \sum_k{\omega_k a_k^\dagger a_k + \frac{\alpha_k}{2} a_k^{\dagger2} a_k^2} \\ 
H_\mathrm{c} &= \sum_{k, l>k}{g_{kl}\left(a_k^\dagger + a_k\right)\left(a_l^\dagger + a_l\right)} 
\label{eq:hamiltonian_4q}
\end{aligned}
\end{equation}
where we consider an all-to-all coupling to account for parasitic couplings. $\omega_k$ and $\alpha_k$ are the qubits' uncoupled frequencies and anharmonicities respectively and $g_{kl}$ is the coupling between qubits $k$ and $l$. 

From this Hamiltonian we can perform both dispersive coupling and rotating wave approximations~\cite{blais_circuit_2021} to get:
\begin{equation}
\begin{aligned}
\tilde H_\mathrm{q} &= \sum_k{\tilde\omega_k \tilde a_k^\dagger \tilde a_k + \frac{\tilde\alpha_k}{2} \tilde a_k^{\dagger2} \tilde a_k^2} \\ 
\tilde H_\mathrm{c} &= \sum_{k, l>k}{\zeta_{kl}\tilde a_k^\dagger \tilde a_k \tilde a_l^\dagger \tilde a_l}
\label{eq:hamiltonian_4q_disp}
\end{aligned}
\end{equation}
where $\tilde\omega_k$ and $\tilde \alpha_k$ are respectively the dressed frequency and anharmonicity of qubit $k$ and $\zeta_{kl}$ is the ZZ interaction between qubits $k$ and $l$.

To estimate the dressed frequencies we use a Ramsey sequence for all individual qubits on the $|g\rangle$-$|e\rangle$ subspace. We estimate the dressed anharmonicity using the same sequence on the $|e\rangle$-$|f\rangle$ subspace and obtain $\tilde{\alpha} = \omega_\mathrm{ef} - \omega_\mathrm{ge}$. To estimate the ZZ interaction $\zeta_{kl}$ we apply a $\pi$ pulse on the qubit $k$ before performing a Ramsey sequence on qubit $l$. Finally, to estimate the bare parameters from the dressed ones, we perform a numerical diagonalization of Eq. \ref{eq:hamiltonian_4q} where the bare parameters are optimized to minimize the difference between the numerically computed and measured dressed parameters (reported in Tables\ref{tab:qubits} and \ref{tab:ZZ}). 

\begin{table}
\caption{\label{tab:qubits}%
\textbf{Qubit parameters}. $\omega_{\mathrm{res}}$ is the readout resonator frequency, $\tilde\omega_{ge}$ is the qubit $g$-$e$ dressed transition frequency, $\tilde \alpha$ is the qubit dressed anharmonicity, $T_1^\mathrm{ge/ef}$ is the relaxation time of the $g$-$e$/$e$-$f$ subspace, and $T_{2,\mathrm{E}}^\mathrm{ge/ef}$ is the $T_{2}$ echo coherence time of the $g$-$e$/$e$-$f$ subspace. Error bars are the standard errors from the fitted values}

\begin{ruledtabular}
\begin{tabular}{lcccc}
\textrm{Parameters}&
\textrm{Switch}&
\textrm{Input}&
\textrm{Output 1}&
\textrm{Output 2}
\\
\colrule
\\
$\omega_{\mathrm{res}}/2\pi$ (GHz) & 6.810 & 5.796 & 7.702 & 6.971\\
$\tilde \omega_{ge}/2\pi$ (GHz) & 4.109 & 3.448 & 4.761 & 4.379\\
$\tilde \alpha/2\pi$ (GHz) & -0.226 & -0.100 & -0.188 & -0.173\\
$T_1^\mathrm{ge}$ (\textmu s) & 53 $\pm$ 5 & 92$\pm$17  & 61$\pm$3&59$\pm$5\\
$T_1^\mathrm{ef}$ (\textmu s)& - & - & 18$\pm$2& 51$\pm$12\\
$T_{2,\mathrm{E}}^\mathrm{ge}$ (\textmu s)& 58$\pm$4& 52$\pm$5& 54$\pm$4& 81$\pm$6\\
$T_{2,\mathrm{E}}^\mathrm{ef}$ (\textmu s)& - & - &  14$\pm$2&35$\pm$6  \\
\end{tabular}
\end{ruledtabular}
\end{table}

\begin{table}
\caption{\label{tab:ZZ}%
ZZ shifts $\zeta_{kl}/2\pi$ and corresponding couplings $g_{kl}/2\pi$ between pairs of qubits (MHz).
}
\begin{ruledtabular}
\begin{tabular}{lcccc}
\textrm{ZZ / coupling} &
\textrm{Switch}&
\textrm{Input}&
\textrm{Output 1}&
\textrm{Output 2}
\\
\colrule
\\
\textrm{Switch} & - & -5.39/52.4 & -0.14/4.86 & -0.56/2.67 \\
\textrm{Input} & -5.39/52.4 & - & -1.11/55.1 & -1.31/43.6 \\
\textrm{Output 1} & -0.14/4.86 & -1.11/55.1 & - & -0.58/6.76 \\
\textrm{Output 2} & -0.56/2.67 & -1.31/43.6 & -0.58/6.76 & - \\
\end{tabular}
\end{ruledtabular}
\end{table}

\section{c-iSWAP gate calibration \label{app:swap_calib}}

\begin{figure*}
\centering
\includegraphics[width=\textwidth]{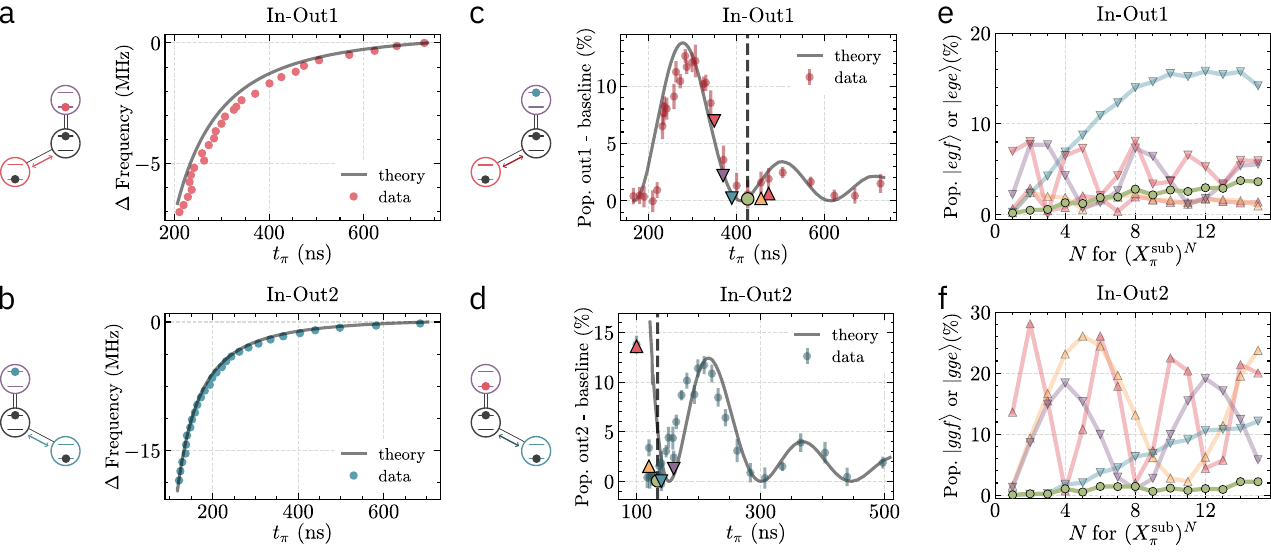}
\caption{\label{figures/fig_SM_2q_calib_minima}
\textbf{C-iSWAP length calibration.}
\textbf{(a, b)} To select the pulse lengths for each iSWAP, we calibrate a series of drive amplitude/frequency tuples corresponding to a selection of different pulse lengths $\tau$ for a $\sqrt{\mathrm{iSWAP}}$, preparing the switch in the state to allow the SWAP to proceed. We find a relationship between $t_\pi\equiv 2\tau$ and the calibrated frequency that matches well with the expected AC-stark shift behavior from simulation of the system Hamiltonian without any fitting parameters.
\textbf{(c, d)} To select a pulse length that minimizes leakage across the two branches of the protocol, for each tuple of calibrated drive amplitude/frequency/length, we prepare the switch in the forbidden state and measure how much population is still transferred to the target output. The observed oscillatory pattern in the transferred population vs. $t_\pi$ matches well with the simulation without fitting parameters (after subtracting off a baseline offset that we attribute to SPAM). We select a length $\tau^*$ for each iSWAP that sits at a minimum of this population transfer (black dashed line) before proceeding to the next layer of finer calibrations (Appendix \ref{app:swap_calib} and Fig. \ref{figures/fig_SM_2q_calib_error_amp}).
\textbf{(e, f)} To quantify the amount of population that is erroneously transferred at $\tau^*$ when the switch is in the blocked state, we perform error amplification as in Fig. \ref{fig3}(g, f) for a few lengths near the minima (selected pulse lengths with their population at $N=1$ highlighted with the corresponding colored marker on (c, d)). The green circles in (e, f) correspond to the populations plotted in Fig. \ref{fig3}(g, f). We include both the $\ket{\bar{S}gf}$ and $\ket{\bar{S}ge}$ states in the population to ensure we capture all the undesired output population, including that which has decayed from $\ket{f}$ to $\ket{e}$ during the error amplification.
}
\end{figure*}

\begin{figure*}[hbt!]
\centering
\includegraphics[width=1\textwidth]{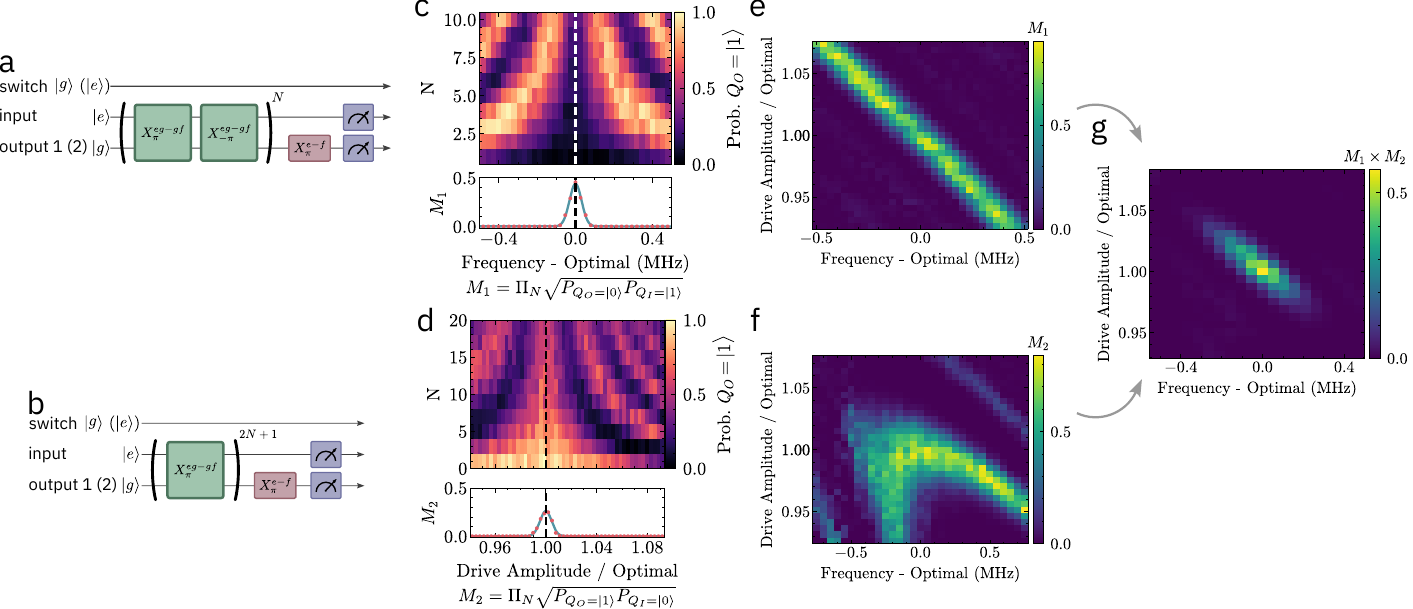}
\caption{\label{figures/fig_SM_2q_calib_error_amp}\textbf{C-iSWAP fine calibration procedure.} \textbf{(a), (b)} Pulse sequences for (c, e) and (d, f) respectively. (a, c, e) represent a $\pi$-minus-$\pi$ experiment, which amplifies frequency error while being robust against amplitude error. (b, d, f) represent a $\pi$-train experiment, which amplifies amplitude error. All data shown is scaled from 0 to 1. \textbf{(c), (d)} As the number of error amplifications $N$ increases, we observe faster and faster oscillations in the frequency (amplitude) that is probed by the $\pi$-minus-$\pi$ ($\pi$-train) experiments. By multiplying the population at each frequency (amplitude) over all $N$, we obtain an easy-to-fit Gaussian that is peaked at the optimal frequency (amplitude). As the number and magnitude of $N$'s sampled increases, the Gaussian peak narrows and the background reduces, allowing for continuous refinement of the optimal point. \textbf{(e), (f)} We perform the $\pi$-minus-$\pi$ ($\pi$-train) experiments at a variety of different amplitudes (frequencies), always fixing the pulse length, so we that we obtain an amplitude vs. frequency map for both the $\pi$-minus-$\pi$ and $\pi$-train experiments. For each amplitude/frequency point, we perform the same multiplication over $N$ as in (c, d) and an additional multiplication over the two qubits measured (input and relevant output). For the $\pi$-minus-$\pi$ experiment we obtain approximately straight lines following the stark shift of the pulse. For the $\pi$-train experiment, we obtain quadratic curves that are symmetric about the stark shift of the pulse. \textbf{(g)} To find the optimal amplitude, frequency tuple for the length we have chosen, we multiply together the results of (e) and (f) to get their intersection, resulting in a blob whose center is the optimal tuple. This tuple is a "stable" point, where alternating updating the frequency from a $\pi$-minus-$\pi$ experiment and updating amplitude from a $\pi$-train experiment will converge to the same amplitude/frequency.}
\end{figure*}

To implement the c-iSWAP gates between input and each of the two outputs, we use flat-topped pulses with Gaussian ramps to calibrate $\sqrt{\mathrm{iSWAP}}$ gates. Two $\sqrt{\mathrm{iSWAP}}$s are tiled to build an iSWAP. There are four possible parameters for each $\sqrt{\mathrm{iSWAP}}$: amplitude $A$, frequency $f$, pulse length $\tau$, and ramp time $\tau_r$, all of which are interdependent. We begin by fixing $\tau_r$ to be 3\,ns each, taking $2\sqrt{2}$ standard deviations for each ramp. We verify using simulations that this ramp time provides a good balance between speed and avoiding unwanted excitations. Importantly, when the AC-stark shift effect is large, having a ramp time that is a sizable fraction of the total pulse length can cause large shifts in optimal frequency when changing the total length~\cite{lu_high-fidelity_2023}. Next, to fix the ($A$, $f$, $\tau$) triplet, we must satisfy two optimization goals. First, for a given $\tau$, we want to find the unique ($A$, $f$) tuple that implements a high fidelity $X^\mathrm{sub}_\pi$ gate where $X_\pi^\mathrm{sub}$ refers to a $\pi$-pulse (iSWAP) in the $\ket{eg}-\ket{gf}$ subspace.
Second, we want to pick the shortest $\tau$ that will not cause leakage between the two branches of c-iSWAPs. To satisfy these two goals, we follow a procedure in which we (1) use rough calibrations to obtain an initial ($A$, $f$, $\tau$) at a range of $\tau$s, (2) check each ($A$, $f$, $\tau$) for leakage across branches of c-iSWAPs to pick an optimal $\tau$, and (3) perform a final fine calibration of ($A$, $f$) at the selected $\tau$.

\begin{enumerate}
    \item \textbf{Rough calibration of $A$ vs $f$ relationship.} We would like to find a series of guess tuples of $A_0$ vs. $f_0$ for a range of $\{\tau_0\}$ that we will sweep. To find the guess tuples ($A_0$, $f_0$), we perform a Rabi chevron experiment sweeping $A$ vs. $f$ at a fixed $\tau$, find $A_0$ that maximizes the swap amplitude at each given $f_0$, and fit the resulting points to a quadratic function. This experiment can be repeated at a few different $\tau$s to ensure the fit is good across the full range of lengths (Figs. \ref{figures/fig_SM_2q_calib_minima}(a, b)). We can then use this quadratic fit to calculate the expected frequency for any given amplitude, matching these ($A_0$, $f_0$) tuples to some $\tau_0$ within the range of tested lengths.
    
    \item \textbf{Looping over tuples: rough calibration of $\tau$.} For each guess tuple ($A_0$, $f_0$, $\tau_0$) where $\tau_0$ is allowed to be off by a factor of two to three, we know that the actual best length $\tau_1$ will fall within a rough range based on the input lengths for the amplitude vs. frequency chevron above. Thus, for every ($A_0$, $f_0$) tuple, we perform a Rabi sweep across a reasonable range of lengths and fit the oscillation to a sinusoid to get an updated length parameter $\tau_1$. At this point, we fix the length parameter to $\tau_1$ for all further calibrations.

    \item \textbf{Looping over tuples: first refinement of $f$.} For each guess tuple ($A_0$, $f_0$, $\tau_1$), we perform a $(X_\pi^\mathrm{sub}, X_{-\pi}^\mathrm{sub})^N$ ("$\pi$-minus-$\pi$") experiment, which finds the AC stark-shifted resonant frequency while being mostly insensitive to amplitude error. In this experiment, we sweep the pulse frequency and an integer number of cycles $N$ (Fig. \ref{figures/fig_SM_2q_calib_error_amp}(a)). As we approach the optimal frequency for the given length and amplitude, we observe slower and slower oscillations in $N$. This $N$ vs. frequency sweep can be fitted using a Gaussian (Fig. \ref{figures/fig_SM_2q_calib_error_amp}(c)) for the optimal frequency $f_1$ by multiplying $\prod_N (1-P_{\ket{1}}(f))$ for the output qubit and $\prod_N (P_{\ket{1}}(f))$ for the input qubit, as in Appendix E3 of~\cite{rower_suppressing_2024}.

    \item \textbf{Looping over tuples: first refinement of $A$.} For each guess tuple ($A_0$, $f_1$, $\tau_1$), we perform a $(X_\pi^\mathrm{sub}, X_{\pi}^\mathrm{sub})^{2N+1}$ ("$\pi$-train", Fig. \ref{figures/fig_SM_2q_calib_error_amp}(b)) experiment, which gives a result that is similar qualitatively to the previous $\pi$-minus-$\pi$ experiment but sweeps the gain instead of the frequency. In the $\pi$-train experiment we choose to perform an odd number of $X_\pi^\mathrm{sub}$ so that a full swap has contrast with the case where no swap occurs at all. The optimal gain $A_1$ is fit to a Gaussian, multiplying $\prod_N (P_{\ket{1}}(f))$ for the output qubit and $\prod_N (1-P_{\ket{1}}(f))$ for the input qubit (Fig. \ref{figures/fig_SM_2q_calib_error_amp}(d)). At this point, we have a set of ($A_1$, $f_1$) tuples for each length $\tau_1$ that are each within 1-2\% of amplitude / a few 100\,kHz from the best final tuple.

    \item \textbf{Optimization for leakage.} We would now like to pick an optimal length $\tau^*$ from our set of calibrated tuples ($A_0$, $f_1$, $\tau_1$) that corresponds to the fastest pulse that enables minimal transfer of population to the wrong output branch. As discussed in Sec. \ref{sec:swap_calib_leakage} of the main text, to calibrate the swap between input and output 1 (2) conditioned on switch in $\ket{g}$ ($\ket{e}$), we prepare the switch in the wrong state $\ket{e}$ ($\ket{g}$) and observe the population transferred to output 1 (2). We pick the optimal length $\tau^*$ to be at the minima with the shortest gate length where the experiments are well-behaved (Figs. \ref{figures/fig_SM_2q_calib_minima}(c, d)). To verify that we have picked an optimal length and upper bound the undesired population transfer, we amplify the measurable population transfer error as in Fig. \ref{fig3}(g, h) for a few different lengths near $\tau^*$ (combined with corresponding calibrated amplitude and frequency) by applying $X_\pi^\mathrm{sub}$ $N$ times with the switch initialized in the wrong state and observe oscillations at lengths away from $\tau^*$ (Fig. \ref{figures/fig_SM_2q_calib_minima}(e, f)).

    \item \textbf{Fine calibration of $A$, $f$.} Once $\tau^*$ is picked, we perform a fine calibration on the amplitude and frequency at $\tau^*$. In our previous calibration, we have assumed that the amplitude obtained from the $\pi$-train experiment and the frequency obtained from the $\pi$-minus-$\pi$ experiment are independent. However, this is not the case due to the AC-Stark shift induced by the drive~\cite{lu_high-fidelity_2023}. We want to make sure that not only is the amplitude correct given the length and frequency, but also the frequency is correct given the length and amplitude. Thus, we perform a fine calibration fixing the length to be $\tau^*$ where we first repeat the $\pi$-minus-$\pi$ experiment at different amplitudes (Fig. \ref{figures/fig_SM_2q_calib_error_amp}(e)) and then repeat the $\pi$-train experiment at different frequencies (Fig. \ref{figures/fig_SM_2q_calib_error_amp}(f)). We thus obtain two 2D maps, each of which sweeps amplitude vs. frequency. The final optimal tuple of $(\tau^*, A^*, f^*)$ is found where the optimal regions in the two 2D maps intersect. We find this tuple by finding the center of the blob that appears when multiplying the two maps together (Fig. \ref{figures/fig_SM_2q_calib_error_amp}(g)).
\end{enumerate}

\section{Single qubit gate calibration \label{app:single_qubit_gates}}
Due to the strong, always-on, ZZ interaction between the qubits, implementing single qubit gates can be challenging. To circumvent this we use the commercial optimal control software Q-CTRL to find pulses that are less sensitive to frequency shifts. To do so, we use the dispersive Hamiltonian from Eq. \ref{eq:hamiltonian_4q_disp} in the frame rotating at $\tilde\omega_k$ for each individual qubit. Then for each qubit $k$ we search for the best possible $X_{\pi/2}$ unitary given the control operator
\begin{equation}
    H_{\mathrm{drive}, k} = \left(I_k(t) +i Q_k(t) \right)\tilde a_k + \textrm{h.c.}
\end{equation}
where the time-dependent coefficients $I_k$ and $Q_k$ are optimized by the solver for a fixed total pulse length that we input. We then vary the pulse length and pick the one that minimizes the infidelity. For the switch, input, output 1, and output 2, we find $X_{\pi/2}$ pulses of lengths $9$, $20$, $13$, and $13$\,ns respectively.
Once the optimal pulse shapes are found we then calibrate them experimentally using the following process:

\begin{enumerate}
    \item\textbf{Ramsey experiment.} To find the pulse frequency we first perform a Ramsey experiment to estimate the dressed qubit frequency.
    \item \textbf{Rabi oscillation versus amplitude.} Using the found qubit frequency and computed pulse shape, we measure the Rabi oscillations versus pulse strength, fit to a sinusoid, and take the drive strength at the maximum of the Rabi oscillation. 
    \item \textbf{Calibration of the virtual Z gate.} The pulses are designed to be optimal in the qubit frequency frame. However, we usually find experimentally that the rotation axis of the implemented pulse has a small tilt toward the Z axis. We attribute this tilt to an AC-stark shift that is not perfectly captured by the simulated evolution. To correct for this phase accumulation, we add a virtual Z phase at the end of each X/2 rotation. To calibrate the virtual Z rotation we perform a $(X_{\pi/2}, Z_\phi, X_{-\pi/2}, Z_{\phi})^N$ sequence where the $Z_\phi$ gate is implemented virtually by updating the carrier phase of the next physical pulse~\cite{mckay_efficient_2017}. We then measure the qubit population versus both $N$ and $\phi$. This experiment is similar to the $\pi$-minus-$\pi$ experiment in Appendix \ref{app:swap_calib} (Fig. \ref{figures/fig_SM_2q_calib_error_amp}(c)), updating the phase instead of sweeping the pulse frequency, and we can similarly fit the product over $N$ to a Gaussian to find the optimal phase correction. 
    \item \textbf{Error amplification versus pulse amplitude.} Once the pulse frequency and virtual Z gate angle are known we perform a $(X_{\pi/2}, X_{\pi/2})^{2N}$ experiment and fit the product over $N$ to a Gaussian to accurately estimate the optimal pulse amplitude, similar to the $\pi$-train experiment in Appendix \ref{app:swap_calib} (Fig. \ref{figures/fig_SM_2q_calib_error_amp}(d)).
\end{enumerate}
We repeat these steps to get a unique frequency, drive amplitude, and virtual Z for each ZZ-shifted pulse. Doing so allows us to prepare higher fidelity product states when more than one qubit is excited.


All of the $e$-$f$ transitions are implemented with simple Gaussians where we take a total of $4\sqrt{2}$ standard deviations per $\pi/2$ pulse. Only the output qubits have $e$-$f$ transitions that are relevant to the protocol performance; these gates are 60\,ns long each. We perform the calibrations for these pulses in a similar manner as for the optimal control-based $g$-$e$ pulses, except that we do not need calibrate the virtual Z corrections. 
We also ensure that the $e$-$f$ pulses are calibrated for the correct ZZ-shifted frequency whenever possible. Note that within the protocol, the $e$-$f$ pulse on output 1 will only ever expect to see the switch in $\ket{g}$, and the $e$-$f$ pulse on output 2 will only ever expect to see the switch in $\ket{e}$, so they do not need to be robust against frequency shifts.

Finally, we implement all $Y_{\pi/2}$ rotations by applying the calibrated $X_{\pi/2}$ pulse with the carrier phase shifted by $\pi/2$. The $X_{\pi}$ and $Y_{\pi}$ are constructed by repeating the respective $\pi/2$ pulse.

\section{Subspace randomized benchmarking of c-iSWAP gates \label{app:rb}}


In this section we present the methods used to extract the average error per swap as well as the leakage out of the logical subspace of the c-iSWAPs.

\subsection{Randomized benchmarking experiment}
We perform two rounds of benchmarking, a reference standard randomized benchmarking experiment and an interleaved benchmarking experiment. We treat the states $\ket{Seg}-\ket{Sgf}$ as the two level logical subspace of a single dual-rail qubit, where $\ket{S}$ is the switch state that allows the SWAP to proceed. For each round of benchmarking, we initialize the system in $\ket{Seg}$ then apply a randomly selected sequence of gates from the 24 unique gates in the Clifford gate set. Each Clifford gate is constructed by tiling 0 (identity), 1, 2, or 3 $X_{\pi/2}^{\mathrm{sub}}$ gates with phase offsets to implement rotations about different axes. At the end of each sequence, we append an extra inversion Clifford. To implement the inversion Clifford, we first calculate the product of all previous gates. If this product gate brings the $+Z$ axis back to $+Z$, we simply take the inversion gate as the identity gate, since we only need to measure the counts along the $Z$-axis. Otherwise, we apply the calculated product gate with a negated phase, so that the ideal final state is $\ket{Seg}$. For each sequence depth we generate 20 variations to obtain statistics.


For each gate sequence, we would like to monitor the population of 12 total states, consisting of states within the subspace ($\ket{Seg}$, $\ket{Sgf}$) and leakage states outside of the subspace ($\ket{\bar{S}eg}$, $\ket{\bar{S}gf}$, $\ket{*gg}$, $\ket{*ge}$, $\ket{*ee}$, $\ket{*ef}$). Here, $\ket{\bar{S}}$ is the switch state that forbids the SWAP and $\ket{*}$ means both the $\ket{g}$ and $\ket{e}$ states on the switch. To distinguish these states, we perform two experiments with different readout parameters for every unique gate sequence that we construct. In the first experiment, we measure the switch, input, and target output qubits using readout parameters optimized to distinguish the $\ket{g}$ and $\ket{e}$ states of each qubit. In the second experiment, we re-run the gate sequence and again measure the switch, input, and target output but change the readout parameters for just the output to be optimized to distinguish between $\ket{e}$ and $\ket{f}$. In both cases, we bin each shot into $\ket{g}$ vs. $\ket{e/f}$ or $\ket{g/e}$ vs. $\ket{f}$. We require this two-step readout because we have a poor readout fidelity if we attempt to use a readout point that can simultaneously distinguish $\ket{g}$, $\ket{e}$, and $\ket{f}$.

To correct for readout error, we follow a modified version of the standard confusion matrix correction (i.e. in Appendix \ref{app:tomography_methods_conf_mat}). In particular, the number of bins for the measured counts is not equal to the number of possible final bins that we want to correct the counts into. To obtain the confusion matrix, we prepare each of the $12$ possible states that we want to bin our final counts into and measure them twice using the two-step readout. From this procedure we obtain a confusion matrix $M$ of size 12 rows $\times (8 + 8)$ columns, where the first and second set of 8 columns correspond to the results of the first and second readout experiments, respectively. Despite the confusion matrix being a rectangular matrix, we can still apply the confusion matrix to each measured set of $8+8$ counts and use the methods described in Appendix \ref{app:tomography_methods_readout_error_correction} to find the most likely set of true measured counts binned into the 12 target final states.

\subsection{Error extraction}

To characterize the fidelity of a single SWAP, we consider the behavior of the probability $p_{\ket{Seg}}$ of returning to the target state, $\ket{Seg}$, after the sequence of gates. This probability should follow a double exponential as a function of the number of Clifford gates $N$~\cite{wood_quantification_2018}, which is composed of the leakage out of the subspace measured by the parameter $\lambda_1$, the logical error within the subspace measured by the parameter $\lambda_2$, and the seepage back into the system measured by the parameter $A_0$:
\begin{equation}\label{eq:prob_return}
    p_{\ket{Seg}} = A_0 + B_0 \lambda_1^N + C_0 \lambda_2^N.
\end{equation}
In our case, the dominating errors are due to leakage triggered by a photon loss in one of the underlying qubits. Hence, we have $\lambda_1 \ll \lambda_2 < 1$ and this double exponential is difficult to fit directly. Instead, we restrict ourselves to the return probability $p_{\ket{Seg}}$ post-selected on the probability to stay within the logical subspace $\ket{Seg}-\ket{Sgf}$, given by $p_\mathrm{survival}=p_{\ket{Seg}}/p_\mathrm{sub}$. The probability to stay within the logical subspace follows a single exponential behavior:
\begin{equation}\label{eq:prob_subspace_fit}
p_{\text{sub}}=A + B\lambda_1^N,
\end{equation}
Since seepage is negligible $p_{\text{sub}}$ and $p_{\ket{eg}}$ will go to zero at large N and hence we have $A\sim A_0\sim 0$. Then the post-selected survival probability is a single exponential that is easier to fit:
\begin{equation}\label{eq:prob_survival_fit}
    p_\mathrm{survival} = C + D\left(\frac{\lambda_2}{\lambda_1}\right)^N; 
\end{equation}
for a more rigorous derivation see~\cite{chen_randomized_2025}. Thus, for each benchmarking round, we first fit Eq. \ref{eq:prob_subspace_fit} to obtain the leakage parameters $A$ and $\lambda_1$. We can estimate the average photon loss (or equivalently leakage) per Clifford or interleaved Clifford, respectively $L_\text{ref}$ and $L_\text{int}$, as~\cite{wood_quantification_2018} :
\begin{equation}
L = (1 - A)(1 - \lambda_1).
\end{equation}
Second, we fit Eq. \ref{eq:prob_survival_fit} to obtain $\lambda_2$ since $\lambda_1$ is known for the previous fit. Using these two fits, we can estimate the average errors $\epsilon_\text{ref}$ or $\epsilon_\text{int}$ as in~\cite{wood_quantification_2018}:
\begin{equation}
    \epsilon= 1 - \frac{1}{d}[(d-1)\lambda_ 2 + 1 - L],
\end{equation}
where $d$ is the subspace dimension ($d=2$ in our case). 
Finally, using the error and photon loss rates from both the reference and interleaved benchmarking, we can estimate the photon loss and error per $X_\pi$ swap as~\cite{rol_fast_2019}:
\begin{equation}
\begin{aligned}
    \epsilon_X &= 1 - \frac{1 - \epsilon_\mathrm{int}}{1 - \epsilon_\mathrm{ref}} \\
    L_X &= 1 - \frac{1 - L_\mathrm{int}}{1 - L_\mathrm{ref}}
\end{aligned}
\end{equation}

We note that though we have benchmarked our system like a dual-rail qubit, unlike in a dual-rail qubit we do not treat leakage errors as erasures since they are not detected, so this error is included in the total errors.

\subsection{Randomized benchmarking for alternative working points}\label{app:rb_other_minima}

\begin{figure}
\centering
\includegraphics[width=1\columnwidth]{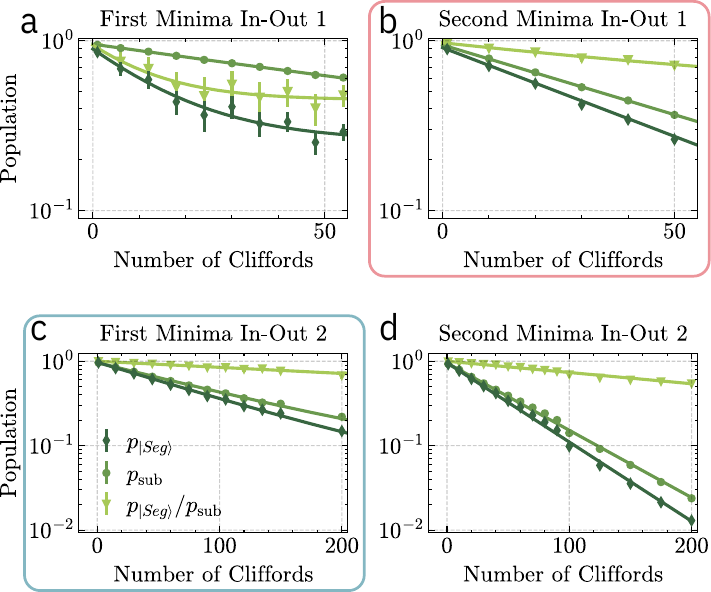}
\caption{\label{fig:rb_bad_minima}\textbf{(a, b)} RB for the iSWAP between input and output 1 for $n=1$ and $n=2$. \textbf{(c, d)} RB for the iSWAP between input and output 2 for $n=1$ and $n=2$. The RB plots corresponding to the selected working points for input to output 1 (2) are boxed in red (blue) and are the same as those shown in Fig. \ref{fig3}(c, d)}
\end{figure}

In Sec. \ref{sec:swap_calib_leakage} and Appendix \ref{app:swap_calib} we saw that several pulse amplitudes can perform a good c-iSWAP, as long as the off-resonant drive rate respects $\Omega = 2ng_\mathrm{eg-gf}$. These points correspond to a minimum of population transfer to the target output qubit when the switch is in the blocking state. If the gates are decoherence-limited, we expect the $n=1$ point to have the best fidelity. Hence, we perform randomized benchmarking on both the $n=1$ and $n=2$ points for the iSWAP between output 1 and 2 (see Fig.\ref{fig:rb_bad_minima}) to pick the best parameter set. 

For the iSWAP between input and output 1, the photon loss rate is $\leakageRbQOneQTwoWrong$ and $\leakageRbQOneQTwo$ for respectively $n=1$ and $n=2$. However, the total errors are, respectively, $\errRbQOneQTwoWrong$ and $\errRbQOneQTwo$. We hypothesize that the higher error rate for the shorter gate is due to a drop in the $\ket{eg}-\ket{gf}$ subspace coherence at larger drive strengths, as has been observed in several experiments~\cite{lu_high-fidelity_2023, nguyen_programmable_2024}. As a consequence, the fastest gate has a lower overall fidelity despite a better photon loss rate, so we use $n=2$ in the protocol. 

On the other hand, for the iSWAP between input and output 2 we have a photon loss rate of $\leakageRbQOneQThree$ and $\leakageRbQOneQThreeWrong$ for respectively $n=1$ and $n=2$. In addition, the total error rates are respectively $\errRbQOneQThree$ and $\errRbQOneQThreeWrong$. For this gate, both the leakage and error rates improve for the shorter gate, indicating that this gate remains limited by photon loss error even at the larger drive strength. Thus, for the protocol we use the $n=1$ working point. 

\section{State tomography}
\subsection{Tomography methods\label{app:tomography_methods}}

In this work, we perform full quantum state tomography on either two or three qubits, which use the same procedure that can be adjusted for $N$ qubits. The procedure can be broken down into five stages: (1) confusion matrix measurements, (2) Pauli basis target state measurements, (3) readout error correction, (4) guess density matrix construction from counts, (5) maximum likelihood estimation (MLE) to find the closest real density matrix from the unphysical guess.

\subsubsection{Measurement of confusion matrix}\label{app:tomography_methods_conf_mat}
To correct for readout error, we apply a standard confusion matrix correction. We define the measurement basis as the $2^N$ bins composed of all combinations of $\ket{g}$ and $\ket{e}$ on $N$ qubits. All measurements in the tomography experiment are made via simultaneous readout tones on all $N$ qubits. For each qubit, the raw demodulated I and Q values are rotated and binned via a fixed threshold into $\ket{g}$ or $\ket{e}$. The rotation angle and threshold are picked to maximize the fidelity between $\ket{g}$ and $\ket{e}$ for that qubit when all other qubits are in $\ket{g}$. We sort each shot into one of the $2^N$ bins by looking at thresholded states on all qubits. To obtain the $2^N \times 2^N$ confusion matrix $M$, we prepare each state in the measurement basis and sort shots into the measurement basis bins. The rows of $M$ correspond to each prepared state and the columns correspond to the measured counts binned into each basis state.

\subsubsection{Measurement of target state in Pauli basis}
Assuming perfect measurements, standard state tomography would require $3^N - 1$ measurements~\cite{nielsen_quantum_2010}. We perform an overcomplete set of $3^N$ measurements corresponding to combinations of measurements along the $X$, $Y$, and $Z$ axes for each qubit. For each of these $3^N$ measurements, we measure all qubits simultaneously and bin individual qubit shots using the same angle and threshold as found during the calibration stage above, then bin collective shots into the measurement basis.

Since all readout tones measure qubits along the $Z$ axis, for the $X$ ($Y$) measurement we perform a $Y_{-\pi/2}$ ($X_{\pi/2}$) "tomography pulse" to align the $\pm 1$ eigenstates of $X$ ($Y$) to the same eigenstate along $Z$. Note that in the presence of ZZ coupling, depending on the initial state and the measurement basis we are considering, the tomography pulses may be quite imperfect since the pulses must be agnostic to the initial state. This issue is addressed further in Sec. \ref{app:ZZ_correction}.

\subsubsection{Readout error correction}\label{app:tomography_methods_readout_error_correction}
After completing all measurements, we move to the post-processing stages of the tomography. We start by correcting for readout error applying the measured confusion matrix on each Pauli basis measurement. Because a direct application of the inverse confusion matrix can sometimes result in negative counts, we use an optimization method to perform the correction as done in QISKIT~\cite{aleksandrowicz_qiskit_2019}. We use the SLSQP optimization method implemented in the publicly available Python package \texttt{scipy} to minimize the cost function $f(\vec{p})=\norm{M^t \vec{p} - \vec{p}_0}$, where $\vec{p}_0$ is the column vector of measured counts for one Pauli basis scaled by the total number of counts to produce a probability, and $\vec{p}$ is the column vector of minimized probabilities. The minimization is performed with constraints for norm ($\norm{\vec{p}}=1$) and range (all $p_i \in [0, 1]$).

\subsubsection{Density matrix guess from corrected counts}
Using only the readout error-corrected counts going forward, we construct an initial guess density matrix $\rho_0$ that will later be used as a starting point for the MLE. To do so we construct a tensor $T$ where each $T_{i_1, i_2, ..., i_N}$ is calculated by taking the expectation value of that Pauli matrix when the state is measured in the corresponding basis. Because all measurements are performed in the $Z$ basis up to applying a rotation just before, $T$ values corresponding to the $Z$ and $I$ basis use the same measurement results with an additional minus sign to correct for the difference in eigenvalues between $Z$ and $I$. The guess $\rho_0$ is then formed by summing over tensor products over the 4 Pauli matrices weighted by the corresponding component of $T$:
\begin{equation}
    \rho_0 = \frac{1}{2^N} \sum_{i_1, ..., i_N \in \{X, Y, Z, I\}} T_{i_1, ..., i_N} \sigma_{i_1}\otimes ... \otimes \sigma_{i_N}
\end{equation}

\subsubsection{Maximum likelihood estimation and ZZ correction}\label{app:ZZ_correction}

\noindent\begin{minipage}{\columnwidth}
\begin{algorithm}[H]
\caption{Efficient MLE for tomography (from~\cite{smolin_efficient_2012})}\label{alg:analytical_mle}
\begin{algorithmic}
\State Define $E=\{E_i\}$ as the set of eigenvalues of $\rho_0$, ordered from largest to smallest
\State Define $\{\ket{E_i}\}$ as the corresponding set of normalized eigenvectors of $\rho_0$
\State Define $d = 2^N$
\State Initialize $E'$, the eigenvalues for $\rho_\mathrm{MLE}$, as $E' = \{E'_i\}=0\,\forall i$
\State Initialize an accumulator $a=0$ and a stop counter $i=d$
\While{$E_{i-1} + a/i < 0$}
    \State $a\gets a + E_{i-1}$
    \State $i\gets i-1$
\EndWhile
\ForAll{$j\in \{0, 1, ... i-1\}$}
    \State $E'_j\gets E_j + a/i$
\EndFor
\State Initialize $\rho_\mathrm{MLE}$ in the shape of $\rho_0$ with all zeros
\ForAll{$j\in \{0, 1, ... i-1\}$}
    \State $\rho_\mathrm{MLE}\gets \rho_\mathrm{MLE} + E'_j\ketbra{E_j}{E_j}$
\EndFor

\noindent\Return $\rho_\mathrm{MLE}$
\end{algorithmic}
\end{algorithm}
\vspace{0.25em}
\end{minipage}

Perfectly matching the guess density matrix to the true density matrix would require an infinite number of measurements~\cite{smolin_efficient_2012}. Thus, we need to use MLE to find the density matrix that is most likely to have produced the set of measurements that we have. We use an efficient MLE algorithm from \cite{smolin_efficient_2012}, whose pseudocode we reproduce in Algorithm \ref{alg:analytical_mle}, to perform the MLE in on average less than $0.3$\,s for three qubit tomography. The algorithm requires that we have (1) an orthonormal Hermitian operator basis $\{\sigma_i\}_{i=1}^{d^2}$ where $d=2^N$, (2) a set of values $\{m_i\}$ that are the average noisy measured expectation values of each $\sigma_i$, and (3) noise that is Gaussian.

Because of the presence of strong ZZ coupling between the qubits in the tomography, especially the two qubit tomography on the switch and input, we modify the algorithm to correct for ZZ effects. As in \cite{roy_tomography_2021}, we do this by simulating the evolution of each measurement basis state under the tomography pulse waveforms. For each Pauli basis $\sigma_i$ we thus build a corresponding evolution matrix. Instead of treating each of the measured counts as a measurement in the true $\sigma_i$ basis, we treat them as counts in a modified basis $\{\tilde{\sigma}_i\}$ and perform the MLE relative to this basis.

To perform the simulation, we setup the Hamiltonian of the full qubit system according to Eq. \ref{eq:hamiltonian_4q_disp}, using the measured dressed qubit frequencies (with all other qubits in $\ket{g}$) for $\tilde{\omega}_k$, measured dressed anharmonicities for $\tilde{\alpha}_k$, and measured average ZZ shift between qubits $k$ and $l$ $(\omega_{\ket{11}} - \omega_{\ket{01}}) - (\omega_{\ket{10}} - \omega_{\ket{00}})$ for $\tilde{\zeta}_{kl}$. We then schedule the pulses corresponding to the Pauli basis we are considering, with the waveform amplitude and shape defined directly by the generated pulse shape from the optimal control. We note that it is important to match the timing of the simulated pulse with the timing of the true experimental pulse sequence (including idle times), as given our ZZ shift magnitudes, states can accumulate non-trivial phases due to ZZ effects with even a few ns of mismatch. We construct each initial state by looping over all the measurement basis states and defining the corresponding Fock basis state, then finding the eigenstate of the Hamiltonian that has the maximum overlap with the Fock basis state to build the dressed basis state. After sending each initial state through the pulse sequence, we take the partial trace over the result to reduce it to just the two lowest levels corresponding to the qubits that we are doing tomography on. Finally, from the set of all evolved basis states we define the evolution matrix $R'_i$ corresponding to basis $\sigma_i$, with the evolved basis states along the columns.

Having obtained the simulated evolution matrices $\{R'_i\}$ that describe the performed rotations (as opposed to a perfect rotation $\{R_i\}$), we still need to "renormalize" the basis and find an updated set of counts that will be inputted into the MLE. We do this in two steps. First, we must determine the actual basis that the measured counts correspond to. Because we always perform measurements along the $Z$ axis, our experimental data consists of a set of $n_0$ counts marked as the $+1$ eigenstate along the $Z$-axis and $n_1$ counts marked as the $-1$ eigenstate along the $Z$-axis. We must find a map to some other deformed axis such that we can interpret the counts as a set of $n_0$ counts along the $+1$ eigenstate of this new axis and $n_1$ counts along the $-1$ eigenstate of the same axis. Second, because the rotations that were performed may not be orthogonal, we must orthonormalize the axes to satisfy the assumptions of the MLE algorithm.

\begin{figure}
\centering
\includegraphics[width=0.6\columnwidth]{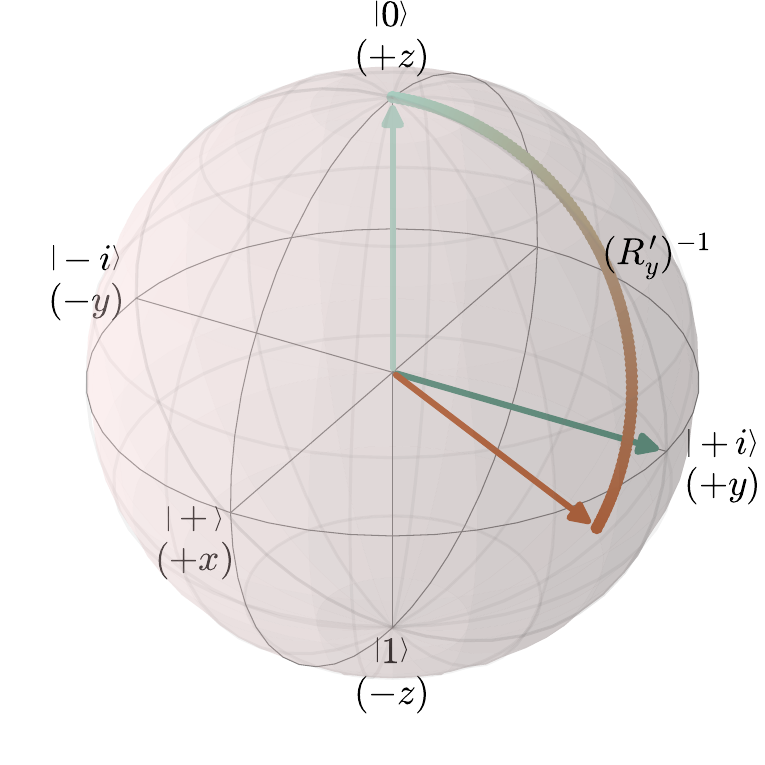}
\caption{\label{figures/fig_SM_zz_correction_traj} During an ideal tomography pulse that corresponds to the target measurement basis $\sigma_i$, the Pauli basis should be rotated to the $+Z$-axis (i.e. dark green state should rotate to light green state; here we show a desired measurement in $\sigma_y$). Instead, under the influence of ZZ interactions, the rotation is deformed to $R_i'$ (here, $R_y'$). In this case, the true axis along which the measured counts can be interpreted is represented by a rotation $(R_y')^{-1}$ from the $+Z$-axis (orange state).}
\end{figure}

We find that the true basis that corresponds to the measured counts is $\{\tilde{\sigma_i}\}=\{(R_i')^\dag \sigma_z R_i'\}$, which can be thought of geometrically as un-rotating the $Z$-axis by the simulated rotation. A diagram of this idea is shown in Fig. \ref{figures/fig_SM_zz_correction_traj}, where we have un-rotated the $Z$-axis by some $R_y'$, a deformed rotation about the $X$-axis that was originally intended as a tomography pulse for the $Y$-basis measurement.

\noindent\begin{minipage}{\columnwidth}
\begin{algorithm}[H]
\caption{Modified Gram-Schmidt for Measurement Basis Orthonormalization}\label{alg:orthonormalization}
\begin{algorithmic}
\State Define $d = 2^n$
\ForAll{$j \in \{0, 1, ..., d-1\}$}
    \State Initialize $\mathcal{N} = \sqrt{\tr(\tilde{\sigma}_j, \tilde{\sigma}_j)}$ \Comment{normalization}
    \State Initialize $\tilde{\sigma}_j^\perp = \tilde{\sigma}_j / \mathcal{N}$
    \State Initialize $p_j^\perp = p_j / \mathcal{N}$
    \ForAll{$k \in \{j+1, j+2, ..., d-1\}$}
        \State $\mathcal{M} = \tr(\tilde{\sigma}_k, \tilde{\sigma}_j^\perp)$ \Comment{calculate overlap}
        \State $\tilde{\sigma}_k = \tilde{\sigma}_k - \mathcal{M} \tilde{\sigma}_j^\perp$ \Comment{subtract parallel component}
        \State $p_k = p_k - \mathcal{M} p_k^\perp$
    \EndFor
    
    \State $\tilde{\sigma}_j^\perp = \sqrt{d}\tilde{\sigma}_j^\perp$
    \State $p_j^\perp = \sqrt{d}p_j^\perp$
\EndFor

\noindent\Return $\{p_j^\perp\}, \{\tilde{\sigma}_j^\perp\}$
\end{algorithmic}
\end{algorithm}
\vspace{0.25em}
\end{minipage}

Next, assuming the ZZ shifts are finite and $\{\tilde{\sigma}_i\}$ completely spans the Hilbert space (i.e. the performed rotations are good enough that no two rotations are exactly the same), we can use Gram-Schmidt decomposition on $\{\tilde{\sigma}_i\}$ and the measured counts to simultaneously orthonormalize the basis and obtain a corresponding set of counts. These renormalized counts can then finally be inputted to the MLE algorithm. For completeness, we provide the orthonormalization algorithm in pseudocode in Algorithm \ref{alg:orthonormalization}, where $\tilde{\sigma}_j$ is already initialized as $(R_j')^\dag \sigma_z R_j'$, and $p_j$ is the $T$ matrix element calculated by treating measurements as made in the $Z$ basis if the target basis measured is $X$, $Y$, or $Z$; or in the $I$ basis if the target basis is $I$.

Note that the final basis matrices $\{\tilde{\sigma}_i^\perp\}$ and counts $\{p_j^\perp\}$ still satisfy the assumption of \cite{smolin_efficient_2012}. The basis is orthonormal as required by the Gram Schmidt. Furthermore the basis remains Hermitian and the noise on the measurement data remains Gaussian (assuming the data started out Gaussian), both because the Gram-Schmidt procedure performs only linear transformations on the data.

\subsection{Virtual Z \label{app:virtual_Z}}

\begin{figure*}[!hbt]
\centering
\includegraphics[width=\textwidth]{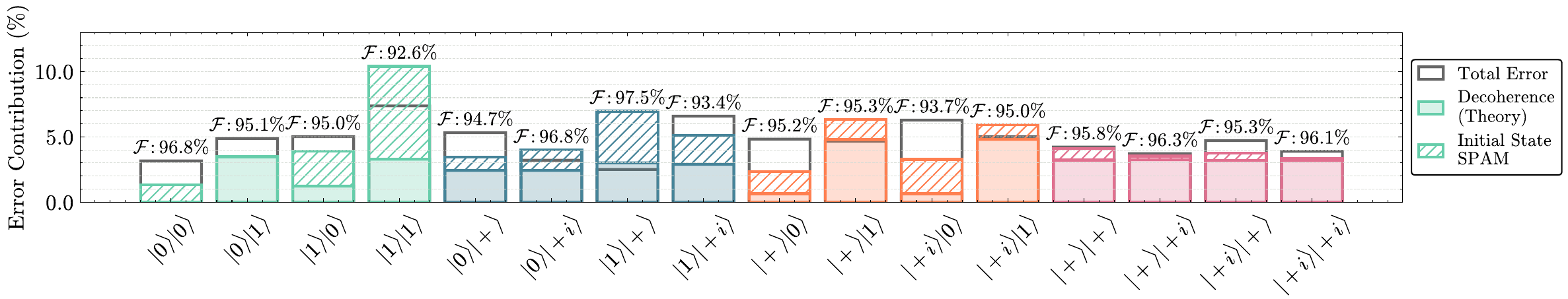}
\caption{\label{fig:app_err_breakdown}\textbf{Error budget for all initial states.}}
\end{figure*}

The routing protocol has the interesting property that it is possible to correct for arbitrary coherent phase errors simply by adding virtual Z gates on the switch and output qubits after each routing operation. In many multi-qubit gate operations, phase errors cannot always be arbitrarily corrected (e.g. a general iSWAP~\cite{ganzhorn_benchmarking_2020}), which at a high level can be explained by there being an insufficient number of knobs to fully correct for the number of independent phases that can be accumulated during the operation. For a general two-qubit iSWAP, we can tune two phases, $\varphi_0$ and $\varphi_1$, by applying a virtual Z gate on each of the two qubits. However, there are three possible phases that can arise during the gate: $\varphi_{\ket{01}}$, $\varphi_{\ket{10}}$, and $\varphi_{\ket{11}}$, corresponding to phase accumulations on each qubit individually and phase accumulation due to the ZZ interaction between the qubits~\cite{ganzhorn_benchmarking_2020}. As a result, one of these phases cannot be corrected.

In the routing case, we also work with two qubits worth of Hilbert space when we consider only the levels that are allowed to have population. Specifically, we can consider the input space basis as four possible $\ket{ab}$, a state on switch/input with output qubits in $\ket{0}$. Similarly, the output space basis consists of four $\ket{abc}$, a state on switch/output 1/output 2 with the input qubit in $\ket{0}$. Thus, within the relevant subspace, just like in the standard two qubit case there are three possible phases that can arise, $\varphi_\alpha$, $\varphi_\beta$, and $\varphi_\gamma$, which correspond respectively to the phases of the $\ketbra{010}{01}$, $\ketbra{100}{10}$, and $\ketbra{101}{11}$ matrix elements relative to $\ketbra{000}{00}$. The unitary for the performed routing operation can be written as
\begin{equation}
    U_{\ketbra{\mathrm{SO_1O_2}}{\mathrm{SI}}}^{\text{route}} =
    \begin{blockarray}{cccccc}
        &&\bra{00} & \bra{01} & \bra{10} & \bra{11}\\
        \begin{block}{cc(cccc)}
        \ket{000} && 1 & 0 & 0 & 0 \\
        \ket{010} && 0 & -ie^{-i\varphi_\alpha} & 0 & 0 \\
        \ket{100} && 0 & 0 & e^{-i\varphi_\beta} & 0 \\
        \ket{101} && 0 & 0 & 0 & -ie^{-i\varphi_\gamma}\\
        \end{block}
    \end{blockarray}
\end{equation}
while the desired unitary $U_{\ketbra{\mathrm{SO_1O_2}}{\mathrm{SI}}}^{\text{route, ideal}}$ is
\begin{equation}\label{eq:U_route}
    U_{\ketbra{\mathrm{SO_1O_2}}{\mathrm{SI}}}^{\text{route, ideal}} = 
    \begin{blockarray}{cccccc}
        &&\bra{00} & \bra{01} & \bra{10} & \bra{11}\\
        \begin{block}{cc(cccc)}
        \ket{000} && 1 & 0 & 0 & 0 \\
        \ket{010} && 0 & -i & 0 & 0 \\
        \ket{100} && 0 & 0 & 1 & 0 \\
        \ket{101} && 0 & 0 & 0 & -i \\
        \end{block}
    \end{blockarray}
\end{equation}

Unlike in the standard two qubit iSWAP case, however, we now have three possible phase knobs that we can adjust, on the switch, output 1, and output 2. Thus, we should be able to fully compensate for any coherent phase accumulated using just virtual Z operations on these three qubits. Each virtual Z operation on qubit $q$ can be written as
\begin{equation}
    Z_q = e^{-i\varphi_q/2}
    \begin{pmatrix}
        1 & 0\\
        0 & e^{i\varphi_q}
    \end{pmatrix}.
\end{equation}
Multiplying out the application of virtual Z gates on the switch, output 1, and output 2, we find
\begin{multline}
    Z_\mathrm{S}Z_\mathrm{O_1}Z_\mathrm{O_2}U_{\ketbra{\mathrm{SO_1O_2}}{\mathrm{SI}}}^{\text{route}} = e^{-i\varphi_0}\times\\
    \begin{pmatrix}
        1 & 0 & 0 & 0 \\
        0 & -ie^{-i(\varphi_\alpha - \varphi_\mathrm{O_1})} & 0 & 0 \\
        0 & 0 & e^{-i(\varphi_\beta - \varphi_\mathrm{S})} & 0 \\
        0 & 0 & 0 & -ie^{-i(\varphi_\gamma - \varphi_\mathrm{S} - \varphi_\mathrm{O_2})} \\
    \end{pmatrix},
\end{multline}
where $\varphi_0$ is just a global phase $(\varphi_\mathrm{S}+\varphi_\mathrm{O_1}+\varphi_\mathrm{O_2})/2$ that can be dropped from the analysis. Based on this expression, we are free to pick $\varphi_\mathrm{O_1}=\varphi_\alpha$, $\varphi_\mathrm{S}=\varphi_\beta$, and $\varphi_\mathrm{O_2}=\varphi_\gamma - \varphi_\beta$, which will perfectly cancel the extra phases to get the desired routing unitary $U_{\ketbra{\mathrm{SO_1O_2}}{\mathrm{SI}}}^{\text{route, ideal}}$. We can also pick a different set of $\varphi_\mathrm{S}$, $\varphi_\mathrm{O_1}$, and $\varphi_\mathrm{O_2}$ to obtain any other arbitrary desired phase on this unitary.

Experimentally, different initial states are sensitive to different virtual Z rotations. For example, an ideally prepared $\ket{0+}_\mathrm{SI}$ should route to $(\ket{000}+\ket{010})_\mathrm{SO_1O_2}$, which is only sensitive to the virtual Z on output 1, while an ideally prepared $\ket{++}_\mathrm{SI}$ should route to 
$(\ket{000}-i\ket{010}+\ket{100}-i\ket{101})_\mathrm{S O_1 O_2}$, which is sensitive to all three virtual Zs applied. To ensure that we have implemented a unitary, the set of three virtual Zs applied to each final state must be the same for all initial states.

\begin{figure}[!htb]
\centering
\includegraphics[width=0.95\columnwidth]{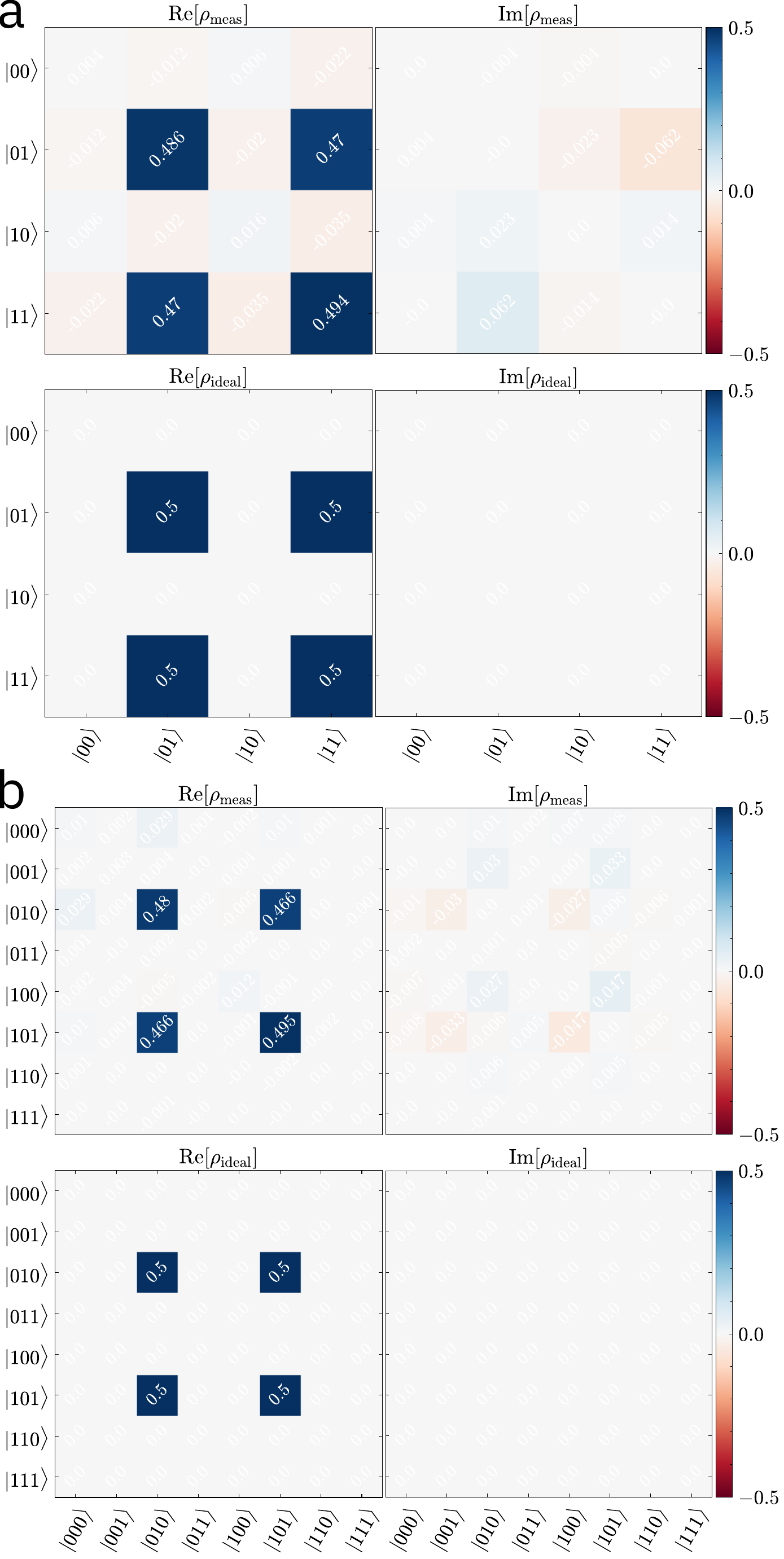}
\caption{\label{fig:app_tomo_comparison}\textbf{Tomography measurements for SPAM.} \textbf{(a)} 2Q tomography on $\ket{+1}_\mathrm{SI}$. In the top (bottom) two panels we plot the measured (ideal) real and imaginary components of the density matrix. \textbf{(b)} 3Q tomography on $(\ket{010} + \ket{101})_{\mathrm{SO}_1\mathrm{O}_2}$, the result of applying the protocol once on the initial state measured in (a). We plot the real and imaginary components of the measured density matrix (top row) and of the target density matrix obtained by applying a perfect protocol unitary on the measured initial state from (a) (bottom row).}
\end{figure}

To test the extent to which this is true, we sweep over the virtual Z applied to the switch, output 1, and output 2 on the measured final state 3Q tomography and find the tuple of phases that obtains the maximum fidelity for the final state ("individual optimization"). We then perform a second search where we look for the tuple of phases that obtains the maximum average fidelity when applied to all 16 different initial states ("global optimization"). For a perfect unitary, the global and individual optimizations should return the same phase tuples. We ascribe any differences to timing errors, where a small discrepancy on the order of a few ns can cause noticeable accumulation of phases due to the large ZZ between the switch and input. In particular, two relevant timescales are the delay between the initial state preparation and the tomography pulses to measure the initial state, and the possibly different delay between the preparation of the initial state and the start of the SWAP gates. The first timescale determines the phase accumulation on the state measured by the initial state tomography (e.g. Fig. \ref{fig:app_tomo_comparison}) and is assumed to be 0 when we perform fidelity calculations referencing an ideal initial state. The second timescale determines the true state seen by the router when it begins the protocol. For different initial states, these delays can be slightly different depending on the exact pulse sequences used to prepare the initial state. The discrepancy between different initial state timings can result in the appearance of the unitary having a phase that is initial state dependent. After optimizing the pulse timing in our setup, we are able to minimize the discrepancy such that we have a remaining fidelity drop of on average $0.4\%$ when using global instead of individual phase optimization. We emphasize that the fidelities quoted in this work use global optimization since our goal is to implement a routing unitary. 

\section{Error contribution for all measured initial states\label{app:err_breakdown}}

In this section, we elaborate on the error breakdown presented in Fig. \ref{fig4}(c). In Fig. \ref{fig:app_err_breakdown} we show the error contribution from the theoretically estimated decoherence and the state preparation SPAM for all 16 cardinal states in which we initialize the router. For all states we use the convention $\ket{+} = \left(\ket{0} + \ket{1}\right)\sqrt{2}$ and $\ket{+i} =\left(\ket{0} + i\ket{1}\right)/\sqrt{2}$. We confirm that the error estimation from decoherence and SPAM closely match the measured ones on average, although there is a state-by-state deviation that we postulate comes mostly from SPAM. Because of the large ZZ interaction between the switch and input, it is difficult to properly prepare product states where one or both qubits have probability of being in $\ket{e}$. Furthermore, tomography pulses are also lower fidelity when measuring these states, and we find that even with ZZ correction (Appendix \ref{app:ZZ_correction}), the reported density matrix does not perfectly recover states that we expect to be prepared with high fidelity. We rely on optimal control pulses (Appendix \ref{app:single_qubit_gates}) to improve the state preparation and tomography measurement, but we find that states such as $\ket{1}\ket{1}$ and $\ket{1}\ket{+}$ are still difficult to prepare and/or to measure with high fidelity.

We note, however, that for the protocol the preparation of the initial state is not relevant to the actual fidelity of the gate. Indeed, as long as the initial state of the system has no population in either of the output qubits, the protocol should be able to correctly route from the switch/input to the switch/outputs. Accordingly, we find that when we compare the measured 3Q density matrix to the measured initial state instead of to an ideal initial state, the measured fidelity improves by $\errSPAMtomo$ on average. In Fig. \ref{fig:app_tomo_comparison}(a), we show the real and imaginary components of the measured preparation of $\ket{+}\ket{1}$ (top row) and of the ideal preparation (bottom row). In Fig. \ref{fig:app_tomo_comparison}(b), we compare the measured 3Q density matrix after sending the $\ket{+}\ket{1}$ state through the protocol to the 3Q density matrix generated by sending the measured initial state through an ideal protocol. The fidelity is then given by  $\mathcal{\bar{F}}=\left(\mathrm{tr}\sqrt{\sqrt{\rho_\mathrm{meas}}\rho_\mathrm{ideal}\sqrt{\rho_\mathrm{meas}}}\right)^2$\cite{nielsen_quantum_2010}, where $\rho_\mathrm{ideal}$ is the measured input state sent through ideal protocol and $\rho_\mathrm{meas}$ is the measured final density matrix.

\section{Scaling the router for a QRAM\label{app:scaling}}

\begin{figure*}[!htb]
\centering
\includegraphics[width=\textwidth]{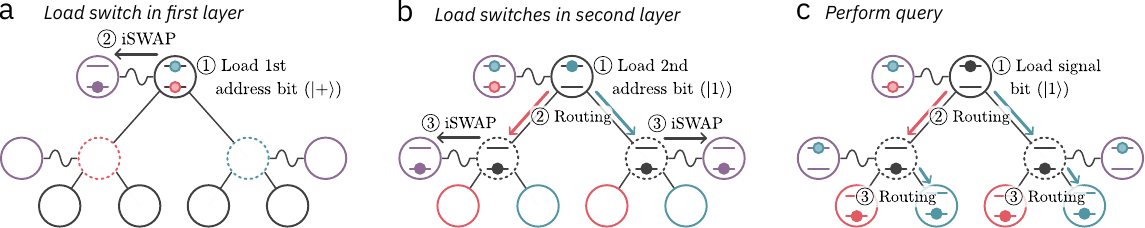}
\caption{\label{figures/scaling} \textbf{Querying a two-address-bit QRAM.} \textbf{(a)} The first input qubit is in gray and the output qubits of the first layer, which will later act as the inputs of the second layer, are in red and blue. In this architecture, we route the input excitation in the tree using alternatively its $\ket{e}$ (solid circle) or $\ket{f}$ (dashed circle) states. Each input/output qubit in the tree is coupled to its corresponding switch (purple) via a tunable coupler, (S-shaped line). Before performing a query, we load the switches one layer at a time as in the bucket-brigade algorithm for QRAM operation. To load the switches of both layers of the tree, we begin by loading the input with the first address state (here, $\ket{+}$) and transferring the state to the switch with an iSWAP (black arrow). \textbf{(b)} Next, we load the switches of the second layer. We initialize the input with the second address bit (here, $\ket{1}$) and use our Q$^2$-router protocol to route this state to some superposition of the two outputs defined by the previously loaded first switch state ($\ket{+}$). We then perform an iSWAP between each second-layer switch and its corresponding input qubit. Since each Q$^2$-router sub-module in the tree is essentially independent, these iSWAPs can be performed simultaneously to ensure that the gate time is constant as a function of the layer depth. \textbf{(c)} Once all the switches are initialized, we can run the query operation using some input state (here $\ket{1}$) by applying the protocol once on each switch/input/output 1/output 2 Q$^2$-router sub-module, starting from the top of the tree. These routing protocols can also be applied simultaneously for each layer.}
\end{figure*}

In this section we propose how the Q$^2$-router we have implemented could be integrated into a QRAM with a few address layers.

A QRAM can be implemented via either a gate-based architecture, which uses a standard QPU and compiles each query of the memory into other logical gates, or a router-based architecture, where queries are implemented via routing operations~\cite{xu_systems_2023}. Compared to the gate-based approach, for the same memory size the router-based solution is exponentially more efficient in the number of sequential operations required for a single query~\cite{xu_systems_2023}. Like a classical RAM fanout architecture, a router-based QRAM has a binary tree architecture where each level $k$ of the tree corresponds to the $k$-th bit of a base-2 memory address of length $n$. The binary tree can specify a total of $2^n$ memory locations. When information reaches a node (switch), it can be routed to the left, right, or a superposition of the two depending on if the switch is in the 0, 1, or a superposition state. To query a single memory location or a superposition of memory locations, we can perform a routing operation from the root node of the tree all the way to the specified memory locations at the leaves of the tree, swap the information from the storage locations back into the tree, and perform the inverse routing operation to bring the retrieved information back to the root node.

To build a scalable QRAM, we propose to operate via the principles of the "bucket-brigade" algorithm~\cite{giovannetti_quantum_2008}. Despite the QRAM architecture having a total number of qubits that is exponential in the total number of address bits $n$, this algorithm ensures that with a router-based QRAM architecture, each query only requires a number of $O(n)$ sequentially applied gates~\cite{giovannetti_architectures_2008, giovannetti_quantum_2008} and also improves noise resilience compared to a standard fanoutarchitecture~\cite{giovannetti_architectures_2008} by requiring a lower degree of entanglement within the tree to maintain the routing capability~\cite{weiss_quantum_2024-1}. In the bucket-brigade architecture, an additional important step is a loading stage in which the switch state should be loaded with the address state prior to beginning the routing. This operation could be done on our router via a SWAP between the input qubit and the switch. In this paper we have simply initialized the switch directly to focus on the routing protocol only.

The routing protocol that we have implemented in this paper can be thought of as a routing operation in a single-address-bit QRAM. To perform a full query operation, in which we retrieve information from a superposition of address locations defined by the initial switch state and multiply the result by the initial input state, we can perform
\begin{equation}
    U_\mathrm{query} = U_\mathrm{route} Z_{\mathrm{correction}}^{\mathrm{S}\mathrm{O}_1\mathrm{O}_2} U_\mathrm{retrieve}U_\mathrm{route}^{-1} Z_{\mathrm{correction}}^{\mathrm{S}\mathrm{O}_1\mathrm{O}_2},
\end{equation}
where $Z_{\mathrm{correction}}^{\mathrm{S}\mathrm{O}_1\mathrm{O}_2}$ is the tuple of virtual Z gates on the switch and outputs that we can calibrate via a single protocol operation (Appendix \ref{app:virtual_Z}) and $U_\mathrm{retrieve}$ is some operation that transfers information from a set of classical or quantum memory storage locations to the corresponding output qubit. For a classical database, for example, $U_\mathrm{retrieve}$ can be a set of classically-controlled $Z$ gates on each final output/classical memory pair, where the control is the value of the classical memory at each address~\cite{hann_resilience_2021}.

To tile our Q$^2$-router into a QRAM with two or more address bits, we propose to use the same main operating principles with a few changes to ensure the bucket-brigade algorithm can be implemented. In Fig. \ref{figures/scaling} we show a schematic of a QRAM with two address bits. We first note that we do not need to use the $g$-$e$ subspace for all qubits in the binary tree. Instead, we can alternate between the $g$-$e$ and $g$-$f$ subspace for even and odd layers in the tree, which will allow the routing to be performed using $\ket{eg}-\ket{gf}$ sideband pulses as in our original Q$^2$-router. In addition, we note that the output qubits of each layer become the input qubits of the following layer.

For the router-based bucket-brigade architecture~\cite{giovannetti_architectures_2008}, there are two main steps: (1) loading the switches with the address bits and (2) routing information between the input and output (and vice-versa). In Fig. \ref{figures/scaling}(a, b) we show how the proposed architecture can be used to load the address bits in the switch. In addition to the native c-iSWAP presented in this article, we need an iSWAP between each input/output qubit and its respective switch. This iSWAP can be implemented from an $\ket{eg}-\ket{gf}$ sideband similar to the c-iSWAP that we have calibrated. The switches are loaded one layer at a time, where for the $k$-th address bit we load the bit in the first input, perform $k-1$ layers of routing operations, and perform an iSWAP between each switch on the $k$-th layer and its corresponding input qubit (Fig. \ref{figures/scaling}(a, b)). A full query operation can be accomplished simply by applying each layer of routing operations sequentially from the top to bottom layer (Fig. \ref{figures/scaling}(c)).

One caveat is that to scale up to multiple memory layers, we need to add a tunable ZZ interaction between each switch and its corresponding input/output qubit. If the ZZ interactions are always-on as in our Q$^2$-router, they would block the c-iSWAPs of the following layer. This behavior can be seen from Fig. \ref{figures/scaling}(c) where once all the address bits have been loaded in the switches, without a tunable ZZ the switches of the second layer would ZZ-shift their corresponding qubits (the inputs of the second layer), such that during the routing procedure, the c-iSWAPs between the input of the first layer and the inputs of the second layer would no longer be resonant. A tunable coupler between each input qubit and its switch, implemented for example by a frequency-tunable transmon, will allow the ZZ interaction to be tuned close to zero in the idle position or activated only when we want to drive the c-iSWAPs of the corresponding layer.


\bibliography{biblio}

\end{document}